\documentclass[a4paper]{article}
\usepackage[table]{xcolor}
\usepackage{titletoc}
\usepackage{pdfpages} 
\usepackage[utf8]{inputenc} 
\usepackage[activeacute,english]{babel} 
\usepackage{amsmath,amssymb,amsfonts} 
\usepackage{braket} 
\usepackage{graphicx} 
\usepackage{bigints} 
\usepackage{bm} 
\usepackage{parskip} 
\usepackage{fancyhdr}
\usepackage{charter}
\usepackage{hyperref}
\usepackage{titlesec}
\usepackage[small]{caption}
\usepackage[a4paper, margin=0.7in]{geometry}
\usepackage{tgadventor}

\usepackage{pgfplots}
\usepackage{comment}
\usepackage{tikz}
\usepackage{cleveref}
\crefformat{equation}{(#2\textcolor{blue}#1#3)}
\crefname{equation}{Eq.}{Eqs.}
\crefname{figure}{Fig.}{Figs.}
\crefname{section}{Section}{Sections}
\crefname{table}{Table}{Tabs.}
\usepackage{subcaption}
\usepackage{dsfont}
\usepackage{bbold}
\usepackage{multirow}
\usepackage{booktabs}
\usepackage{xspace}
\newcommand{\sk}{{\scalebox{0.6}{$K$}}}
\newcommand{\mk}{\scalebox{0.8}{$K$}}
\newcommand{\sn}{{\scalebox{0.6}{$N$}}}

\crefrangeformat{equation}{#3(\textcolor{blue}{#1–\hspace{0.5mm}#2})#4}

\usepackage{etoolbox}
\patchcmd{\ref}{\theequation}{Eq.~(\theequation)}{}{}

\newcommand{\caseCI}{CI\xspace}
\newcommand{\caseCII}{CII\xspace}
\newcommand{\caseCIII}{CIII\xspace}
\newcommand{\caseCIV}{CIV\xspace}

\pgfplotsset{compat=1.18}

\providecommand{\abs}[1]{\lvert#1\rvert}

\DeclareMathAlphabet{\mathpzc}{OT1}{pzc}{m}{it}

\title{\textbf{Confluent supersymmetric algorithm for bilayer graphene}}

\author{Jonathan de la Cruz-Hernandez\thanks{jonathan.delacruz@cinvestav.mx}
\hspace{0.4mm} and 
        David J. Fernández C.\thanks{david.fernandez@cinvestav.mx}}

\date{{\fontsize{10.3pt}{12pt}\selectfont
\textit{Departamento de Física, Cinvestav, A.P. 14-740, 07000 Ciudad de México, Mexico}}}

\begin{document}

\maketitle

\renewcommand{\thefootnote}{\fnsymbol{footnote}} 

\begin{abstract}
   External magnetic field profiles leading to equidistant and partially equidistant bilayer graphene spectra within the tight-binding model are obtained. This is achieved by implementing the integral and differential versions of the second-order confluent algorithm to the harmonic oscillator for arbitrary real factorization energies. Additionally, new Barut-Girardello and Gilmore-Perelomov coherent states for bilayer graphene are derived, for both diagonal and non-diagonal ladder operators. Their time evolution is analyzed, finding temporal stability and cyclic evolution in some cases. This fact is contrasted with the non-cyclic evolution of bilayer graphene coherent states obtained when using two different factorization energies. Likewise, the geometric phase and uncertainty product of the quadratures for the previously obtained coherent states are studied.\\
   
   \noindent\textit{Keywords:} bilayer graphene, supersymmetric quantum mechanics, confluent algorithm, coherent states
   
\end{abstract}

\section{Introduction}
\label{sec:introduction}

The physical properties of graphene in magnetic fields has been a subject of intensive study in recent years, from both experimental and theoretical viewpoints \cite{Katsnelson2012}. Graphene is a two-dimensional material composed of one or more layers of carbon atoms arranged in a hexagonal lattice. This material exhibits unique electronic properties, making it an ideal system for exploring new physical phenomena. The dynamics of charge carriers in monolayer graphene under external magnetic fields can be studied using the \textit{tight-binding} model. This approach accurately describes the quantum behavior of electrons in graphene at low energies, where the Dirac-Weyl equation plays a fundamental role \cite{DiVincenzo1984,Neto2009_EPGraphene}. Moreover, an effective Hamiltonian capturing the electronic interactions at low energies for bilayer graphene can be as well derived \cite{Katsnelson2012,McCann2006,McCann2013}.

Within the aforementioned models, analytical solutions for monolayer graphene have been obtained by employing first-order supersymmetric quantum mechanics (SUSY QM) \cite{Setare2008,Kuru2009,Midya2014}. On the other hand, higher-order SUSY QM has proven to be useful for obtaining exact solutions in diverse physical systems, due to the wide range of options available for designing the spectra of the supersymmetric partner Hamiltonian \cite{Fernandez2004_HigherOrder,Fernandez2019_Trends}. In particular, the second-order SUSY QM can be implemented by using either two different real factorization energies (real case), two identical ones (confluent case) or a pair of complex conjugate factorization energies (complex case). Let us note that, by employing second-order SUSY QM in the real case, analytical solutions for the bilayer graphene effective Hamiltonian have been found in the context of shape-invariant potentials \cite{Fernandez2020_ElectronBilayerGraphen}, as well as for potentials without such a symmetry \cite{Fernandez2021_BGMagneticFields}. 
Furthermore, another solutions have been obtained by applying the integral version of the confluent algorithm, for factorization energies within the spectrum of some exactly solvable auxiliary Hamiltonians \cite{Fernandez2021_BGMagneticFields}.
Additionally, coherent states (CS) for both monolayer and bilayer graphene in a constant homogeneous magnetic field have been derived, using either diagonal ladder operators \cite{DiazBautista2017_GraphenCoherentStates,Fernandez2020_BG_CoherentStates,MNN24,Ba24} or non-diagonal ones \cite{DOCR20,Fernandez2022_GGCoherentStates,DNN21,OD24}. However, due to the non-equidistant nature of the energy spectrum, the time evolution of these states in general  does not exhibit temporal stability. 

In this work we will explore an interesting case arising for the supersymmetric partners of the harmonic oscillator, generated via the second-order confluent algorithm. We will show that it is possible to find position-dependent magnetic fields leading to equidistant spectra for the bilayer graphene effective Hamiltonian. Moreover, we will construct the corresponding bilayer graphene coherent states.

Thus, in \cref{sec:effective_hamiltonian}, we will introduce the effective Hamiltonian for bilayer graphene in external magnetic fields within the tight-binding model. In \cref{sec:confluent_algorithm} we will apply the confluent algorithm to a shifted harmonic oscillator for an arbitrary factorization energy, employing the integral and differential methods. In \cref{sec:confluent_algorithm_bilayer_graphene}, we will use the results of \cref{sec:confluent_algorithm} to identify the spectrum for the bilayer graphene effective Hamiltonian. In \cref{sec:coherent_states_bilayer_graphene}, we will derive the bilayer graphene coherent states by introducing both diagonal and non-diagonal ladder operators. Such coherent states will be classified according to the factorization energy employed in the confluent algorithm. In \cref{sec:time_evolution,sec:commutation_relation}, we will address the time evolution and uncertainty product for the new bilayer graphene coherent states, respectively. Finally, we will present our concluding remarks.

\section{Effective Hamiltonian for bilayer graphene}
\label{sec:effective_hamiltonian}

Let us suppose that the bilayer graphene is placed in an external magnetic field perpendicular to the bilayer surface, the $x-y$ plane, which depends just on $x$. With these assumptions, the magnetic field takes the form $\bm B = B(x) \bm e_z$ while the vector potential, in the Landau gauge, can be expressed as $\bm A = A(x) \bm e_y$. We will consider the charge carriers behavior in bilayer graphene through the tight-binding model. Moreover, following the works \cite{Fernandez2020_ElectronBilayerGraphen}, \cite{Fernandez2021_BGMagneticFields} and \cite{Fernandez2022_GGCoherentStates}, we will take into account an additional term in the effective Hamiltonian that depends only on the $x$ direction, namely $-\hbar^2f(x)\sigma_x$ (where $\sigma_x$ is the Pauli matrix with
ones in the off-diagonal). Thus, the eigenvalue equation to be solved is given by 
\begin{equation}
\label{eq:eigenvalues_equation_BG}
    \mathbb{H}\Psi(x,y)=E\Psi(x,y),
\end{equation}
with $\mathbb{H}$ being the effective Hamiltonian for bilayer graphene placed in the external magnetic field, which is given by
\begin{equation}
    \mathbb{H} = \frac{1}{2m*}
    \begin{pmatrix}
        0 & (\pi_x-i\pi_y)^2 - \hbar^2 f(x)\\[1em]
        (\pi_x+i\pi_y)^2 - \hbar^2 f(x) & 0
    \end{pmatrix},
\end{equation}
where $m^*$ is the effective mass, $\bm \pi = \bm p + \frac{e}{c} \bm A$, and $f$ is an arbitrary function. Due to the homogeneity of the magnetic field along $y$ axis, the momentum in such direction is conserved, the associated operator commutes with $\mathbb{H}$, and thus there exists a common set of eigenstates for both operators. It is straightforward to see that these eigenfunctions can be generally written as
\begin{equation}
\label{eq:eigenfuntions_general}
    \Psi(x,y)=e^{iky}
    \begin{pmatrix}
        \psi^{(2)}(x) \\[0.5em]
        \psi^{(0)}(x)
    \end{pmatrix},
\end{equation}
such that the eigenvalue equation \cref{eq:eigenvalues_equation_BG} leads to the following system of equations
\begin{equation}
\label{eq:coupled_system}
    \begin{aligned}
        L_2^-\psi^{(0)}(x) & =-\varepsilon \, \psi^{(2)}(x) \\[0.5em]
        L_2^+\psi^{(2)}(x) & =-\varepsilon \, \psi^{(0)}(x),
    \end{aligned}\quad,
\end{equation}
where $\varepsilon = 2m^*E/\hbar^2$ and 
\begin{equation}
\label{eq:intertwining_operator}
    \begin{array}{cc}
        \begin{array}{c}
            L_2^- = \dfrac{d^2}{dx^2} + \eta(x)\dfrac{d}{dx} + \gamma(x), \\[1em]
            \eta(x) = 2\left(k+\dfrac{e}{c\hbar}A(x)\right),
        \end{array}
        & 
        \begin{array}{c}
            L_2^+ = \left(L_2^-\right)^\dagger = \dfrac{d^2}{dx^2} - \eta(x)\dfrac{d}{dx} + \gamma(x) - \eta'(x),\\[1em]
            \gamma(x) = \dfrac{1}{4}\left(\eta^2(x) + 2\eta'(x)\right) + f(x),
        \end{array}
    \end{array}
\end{equation}
with $k$ being the wave-number in $y$ direction.
The coupled system of equations \cref{eq:coupled_system} can be easily decoupled by applying $L_2^+$ to the first equation and $L_2^-$ to the second, leading to
\begin{equation}
\label{eq:decoupled_system}
    \begin{aligned}
        L_2^+L_2^-\psi^{(0)}(x) & =\varepsilon^2 \, \psi^{(0)}(x), \\[0.5em]
        L_2^-L_2^+\psi^{(2)}(x) & =\varepsilon^2 \, \psi^{(2)}(x).
    \end{aligned}
\end{equation}
Looking at the previous equations, we realize that the second-order supersymmetric quantum mechanics can be applied to obtain the corresponding solutions. In order to do that, let us consider two auxiliary Hamiltonians
\begin{equation}
    H_j=-\dfrac{d^2}{dx^2}+V_j(x),
\end{equation}
whose eigenfunctions are denoted $\psi_n^{(j)}$, with $j = 0,2$. Let us suppose that both Hamiltonians are intertwined by the operator $L_2^-$ of Eq. \cref{eq:intertwining_operator} in the way
\begin{equation}
\label{eq:operatorial_intertwining}
    H_2L_2^-=L_2^-H_0.
\end{equation}
In order to fulfill this operator relationship, the function $f$ must be given in terms of the function $\eta$ as follows
\begin{equation}
    f(x) 
    = \dfrac{\eta'{}^2(x)-2\eta(x)\eta{''}(x)-
    (\epsilon_1-\epsilon_2)^2}{4\eta^2(x)},
\end{equation}
with $\epsilon_1, \epsilon_2 \in \mathbb{C}$ being two arbitrary constants called factorization energies, which are closely related to the SUSY transformation. Furthermore, the corresponding SUSY partner potentials are given by the following expressions
\begin{equation}
    V_0(x)=-\gamma(x)
    +\dfrac{1}{2}\left(
    \eta^2(x)-\eta'(x)+\epsilon_1+\epsilon_2
    \right),
\end{equation}
\begin{equation}
\label{eq:SUSY_partner_potential}
    V_2(x)=V_0(x)+2\eta'(x).
\end{equation}

The real case with two different real factorization energies, for several \textit{initial} potentials $V_0$, has been studied in various works previously  (see e.g. \cite{Fernandez2020_ElectronBilayerGraphen} and \cite{Fernandez2021_BGMagneticFields}). However, it is also possible to set both factorization energies to be the same, $\epsilon_1 = \epsilon_2 \equiv \epsilon$, a case called \textit{confluent algorithm}, which can be used to generate interesting solutions for bilayer graphene. In this case, the seed solution $u$ associated with the factorization energy $\epsilon$ must be annihilated by the intertwining operator $L_2^-$, and it must be a formal eigenfunction of the initial Hamiltonian $H_0$, i.e., $H_0 u = \epsilon u$. The spectrum of the \textit{new} Hamiltonian $H_2$ depends on the factorization energy $\epsilon$ used and the square integrability of the formal eigenfunction $\psi_\epsilon^{(2)}$ of $H_2$; meanwhile, the other eigenfunctions $\psi_n^{(2)}$ expressed in terms of those of $H_0$, $\psi_n^{(0)}$, are given by
\begin{equation}
\label{eq:eigenfunctions_H2}
    \psi_n^{(2)}(x)
    =\dfrac{L_2^-\psi_n^{(0)}(x)}{\abs{\mathcal{E}_n-\epsilon}},
    \quad \mathcal{E}_n\neq\epsilon,
\end{equation}
where $\mathcal{E}_n$ is the corresponding eigenvalue of $H_0$ and $H_2$. In conclusion, the eigenstates for bilayer graphene are given by Eq. \cref{eq:eigenfuntions_general}, with the pair $\psi_n^{(0)}$, $\psi_n^{(2)}$ as described previously, except by the state $\psi_\epsilon^{(2)}$, as we will see later. Additionally, by considering the intertwining relationship \cref{eq:operatorial_intertwining}, it follows that
\begin{equation}
\label{eq:solution_system}
    \begin{aligned}
        L_2^+L_2^-\psi_n^{(0)}(x)
        &=(H_0-\epsilon)^2\psi_n^{(0)}(x)
        =(\mathcal{E}_n-\epsilon)^2\psi_n^{(0)}(x),\\[0.5em]
        L_2^-L_2^+\psi_n^{(2)}(x)
        &=(H_2-\epsilon)^2\psi_n^{(2)}(x)
        =(\mathcal{E}_n-\epsilon)^2\psi_n^{(2)}(x).
    \end{aligned}
\end{equation}
Thus, taking into account expression \cref{eq:decoupled_system}, the eigenenergies of the electrons in bilayer graphene are given by
\begin{equation}
\label{eq:eigenvalues_BG}
    E_n=\dfrac{\hbar^2}{2m^*}\abs{\mathcal{E}_n-\epsilon}.
\end{equation}

Note that the specific form of the spectrum and eigenstates of $\mathbb{H}$ depends on the solutions of the \textit{auxiliary} Schrödinger problems associated with the intertwined Hamiltonians $H_0$ and $H_2$. Following the framework of SUSY QM, given $V_0$ the initial problem associated with $H_0$ is fixed. Next, a factorization energy is chosen, which characterizes the seed solution $u$ employed and the key function $\eta$. The last function, in turn, defines in a precise way the SUSY partner potential $V_2$, as well as its spectrum and corresponding eigenfunctions. Moreover, taking into account the definition of $\eta$, the applied magnetic field turns out to be 
\begin{equation}
\label{eq:magnetic_field}
    B(x)=A'(x)=\dfrac{c\hbar}{2e}\eta'(x).
\end{equation}
As a result, the external magnetic field, as well as the spectrum and eigenstates of bilayer graphene, depend on the factorization energy chosen and the seed solution employed. It is worth noting that the confluent algorithm was used in \cite{Fernandez2021_BGMagneticFields} for some initial potentials $V_0$, but considering just factorization energies within the spectrum of $H_0$. Moreover, the bilayer graphene coherent states have not been derived in the confluent case, and those constructed in the real case correspond to specific factorization energies, namely, the ground state and the first excited state of $H_0$. Our contribution in this work consists in analyzing the solutions for bilayer graphene by considering the harmonic oscillator as initial potential together with the confluent algorithm for an arbitrary real factorization energy $\epsilon$, as well as the derivation of the associated coherent states for any value of $\epsilon$.

\section{Second-order confluent algorithm}
\label{sec:confluent_algorithm}

Let $V_0$ be the initial potential defined in the interval $(x_l, x_r)$, whose eigenvalues and eigenfunctions are $\mathcal{E}_n$ and $\psi_n^{(0)}$, respectively. Let us choose as factorization energy $\epsilon \in \mathbb{R}$, and $u(x)$ as the seed solution used to implement the second-order confluent algorithm; under these assumptions $\eta$ turns out to be \cite{Fernandez2004_HigherOrder}
\begin{equation}
    \eta(x)
    =-\partial_x \log{w(x)}
    =-\dfrac{\partial_x w(x)}{w(x)},
\end{equation}
while the formal eigenfunction of $H_2$ associated with $\epsilon$, called \textit{missing state}, becomes:
\begin{equation}
\label{eq:missing_state}
    \psi_\epsilon^{(2)}(x) \propto \dfrac{\eta(x)}{u(x)} \propto \dfrac{u(x)}{w(x)},
\end{equation}
where $w$ can be obtained in two equivalent ways: either using the \textit{integral method} or the \textit{differential} one (see \cite{Fernandez2019_Trends} and \cite{Astorga2015_IntegralDifferential} for more details). 

In the integral method $w$ is given by
\begin{equation}
\label{eq:w_integralmethod}
    w(x)=w_0-\int_{x_0}^{x}u^2(t)\;\mathrm{d}t,
\end{equation}
where $x_0$ is a fixed point in the domain of $V_0$, which can be chosen at convenience for simplifying the calculations.
Besides, if $w$ has no nodes then the SUSY partner potential $V_2$ has no singularities in the initial domain, thus we will say that the transformation is non-singular. As noted in \cite{Fernandez2003_ConfluentAlgorithm} (see also \cite{FS05}), such a \textit{regularity condition} can be ensured if we choose a seed solution $u$ such that 
\begin{equation}
\label{cond:left}
    \lim_{x\rightarrow x_l}u(x)=0
    \quad \mathrm{and} \quad
    I_l=\int_{x_l}^{x_0}u^2(t)\;\mathrm{d}t<+\infty,
\end{equation}
or 
\begin{equation}
\label{cond:right}
    \lim_{x\rightarrow x_r}u(x)=0
    \quad \mathrm{and} \quad
    I_r=\int_{x_0}^{x_r} u^2(t)\;\mathrm{d}t<+\infty.
\end{equation}
Taking into account these conditions, $w$ will be nodeless if the parameter $w_0$ belongs to the regularity domain $(-\infty, -I_l]$ or $[I_r, +\infty)$, respectively. Furthermore, if $u$ is square-integrable the two previous conditions will be satisfied, thus the regularity condition is fulfilled if $w_0 \in (-\infty, -I_l] \cup [I_r, +\infty)$.
In the remainder of this work, if $w_0\equiv w_r$ lies inside one regularity interval, we will be dealing with the so-called \textit{regular case}. On the other hand, if $w_0\equiv w_c$ coincides with one extreme value of the domain (either $-I_l$ or $I_r$), we will be working with the \textit{critical case}. 

Note that the spectrum of the new Hamiltonian $H_2$ depends on the square integrability of the missing state \cref{eq:missing_state}. For example, if we choose $\epsilon=\mathcal{E}_j$ and $u(x)=\psi_j^{(0)}(x)$, in the regular case the spectra of $H_2$ and $H_0$ coincide, thus the SUSY transformation is isospectral. However, for another choice of factorization energy with the parameter $w_0$ inside the regularity domain, a new energy level associated with $\epsilon$ is created in the spectrum of $H_2$. Taking into account these considerations, from now on the seed solution will be denoted as $u(\epsilon, x)$ to indicate explicitly its dependence on the factorization energy $\epsilon$.

On the other hand, in the differential method the function $w$ is given by \cite{Bermudez2012,Be2016}
\begin{equation}
\label{eq:w_differential}
    w(\epsilon, x)=W_0+W[u,\partial_\epsilon u](\epsilon, x),
\end{equation}
where $W[u,\partial_\epsilon u](\epsilon, x)
=u(\epsilon, x)\partial_x(\partial_\epsilon u(\epsilon, x))- \partial_x u(\epsilon, x) \partial_\epsilon u(\epsilon, x)$ is the Wronskian in the variable $x$ of $u$ and its parametric derivative $\partial_\epsilon u$. Additionally, it can be shown that  \cite{Astorga2015_IntegralDifferential} (see also \cite{SY18})
\begin{equation}
    \int_{x_0}^{x}u^2(\epsilon; t)\;\mathrm{d}t=
    W[u,\partial_\epsilon u](\epsilon; x_0)
    -W[u,\partial_\epsilon u](\epsilon; x),
\end{equation}
which along with Eq.~\cref{eq:w_integralmethod} implies that
\begin{equation}
    W_0=w_0-W[u,\partial_\epsilon u](\epsilon; x_0).
\end{equation}
Thus, the regularity condition \cref{cond:left} or \cref{cond:right}, in terms of the Wronskian $W[u, \partial_\epsilon u]$, becomes
\begin{equation}
    W_0\in\;(-\infty,-W[u,\partial_\epsilon u](\epsilon; x_l)],
\end{equation}
or
\begin{equation}
    W_0\in[-W[u,\partial_\epsilon u](\epsilon; x_r),+\infty),
\end{equation}
respectively. Moreover, if $u$ is square-integrable both equations \cref{cond:left} and \cref{cond:right} are satisfied, thus in this case
\begin{equation}
    (-\infty,-W[u,\partial_\epsilon u](\epsilon; x_l)]\cup[-W[u,\partial_\epsilon u](\epsilon; x_r),+\infty),
\end{equation}
is the regularity domain of $W_0$ for the SUSY partner potential $V_2$.

\subsection{Confluent algorithm applied to the shifted harmonic oscillator}
\label{subsec:confluent_algorithm_harmonic_oscillator}

Let us consider as initial system a displaced harmonic oscillator potential of the form
\begin{equation}
\label{eq:harmonic_oscillator_potential}
    V_0(x)
    =\dfrac{\omega^2}{4}\left(x+\dfrac{2k}{\omega}\right)^2-\dfrac{\omega}{2},
\end{equation}
with domain of definition being the full real line. The eigenvalues and eigenfunctions of the corresponding stationary Schrödinger equation are given by~\cite{Kuru2009}
\begin{equation}
    \label{eq:solutions_HO}
    \mathcal{E}_n = n\omega, \qquad\psi_n^{(0)}(x) = \left(2^nn!\sqrt{2\pi/\omega}\right)^{-1/2} e^{-z^2/4} H_n\left(\dfrac{z}{\sqrt{2}}\right),
    \qquad n \in \mathbb{N}_0,
\end{equation}
where $z = \sqrt{\omega}(x + 2k/\omega)$ and $H_n$ denotes the Hermite polynomial of degree $n$. Moreover, the general solution to the differential equation $H_0 u = \epsilon u$ for any arbitrary factorization energy $\epsilon$ can be written as follows
\begin{equation}
\label{eq:general_solution}
    u(\epsilon, z) 
    = e^{-{\frac{z^2}{4}}}\Bigg[A
    {}_1F_1\left(-\frac{\nu}{2}, \frac{1}{2}; \frac{z^2}{2}\right)
    + B z
    {}_1F_1\left(\frac{1-\nu}{2}, \frac{3}{2}; \frac{z^2}{2}\right)\Bigg],
\end{equation}
with $A$ and $B$ being arbitrary constants, $\nu = \epsilon / \omega$ and ${}_1F_1$ is the confluent hypergeometric function~\cite{AbraSteg72}. Nevertheless, the solution that satisfies Eq.~\cref{cond:right}, leading to the regularity interval, is the parabolic cylinder function $u(\epsilon, z) = D_\nu(z)$, which is obtained from the previous expression by taking  
\begin{equation}
    A = \dfrac{2^{\frac{\nu}{2}}\sqrt{\pi}}{\Gamma\left(\frac{1-\nu}{2}\right)},\qquad 
    B = -\dfrac{2^{\frac{\nu+1}{2}}\sqrt{\pi}}
    {\Gamma\left(-\frac{\nu}{2}\right)}.
\end{equation}
Meanwhile, the solution fulfilling Eq.~\cref{cond:left} is the previous parabolic cylinder function reflected in the $z$-argument, i.e., $u(\epsilon, -\sqrt{\omega}(x+2k/\omega))$. Both linearly independent solutions are appropriate for implementing the confluent algorithm; in this work, we will use only $u(\epsilon, z)= D_\nu(z)$ for this purpose. This solution vanishes for large values of $z$ according to
\begin{equation}
    D_\nu(z)\sim e^{-z^2/4}z^\nu \quad \mathrm{as} \quad z\rightarrow +\infty, \hspace{3mm} \nu \in \mathbb{R},
\end{equation}
which ensures that, for a given $x_0 \in \mathbb{R}$, the condition \cref{cond:right} is fulfilled for any $\epsilon \in \mathbb{R}$. For simplicity, from now on we will take $x_0 = +\infty$; then
\begin{equation}
    W[u,\partial_\epsilon u](\epsilon, +\infty)
    :=\lim_{x\rightarrow +\infty}W[u,\partial_\epsilon u](\epsilon, x)=0
    \hspace{3mm} \implies \hspace{3mm} W_0=w_0.
\end{equation}
Therefore, the regularity interval for the SUSY partner potential $V_2$ is $w_0 \in [0,+\infty)$.

In particular, for $\epsilon=\mathcal{E}_j$ we will have
\begin{equation}
    D_j(z)=2^{-j/2}e^{-z^2/4}H_j\left(\dfrac{z}{\sqrt{2}}\right)=(2\pi)^{1/4}\sqrt{j!}\;\psi_j^{(0)}(x),
\end{equation}
such that both conditions \cref{cond:left} and \cref{cond:right} are satisfied. Thus, for $w_0 \in \big(-\infty, -\sqrt{2\pi} j! \big] \cup \big[0, +\infty \big)$ we will get a regular SUSY partner potential $V_2$.

Let us consider first the integral method. By plugging the parabolic cylinder function $D_\nu$ into Eq.~\cref{eq:w_integralmethod} it turns out that
\begin{align}
     w(\epsilon,x) &=  w_0 + \frac{2^{\nu-1} z e^{-\frac{z^2}{2}}}{\sqrt{\omega}} \sum_{n=0}^{+\infty} \sum_{k=1}^{n+1}
         \frac{n! z^{2n}}{2^n (n-k+1)!} \left[ \frac{\Gamma(\frac{1}{2})(-\frac{\nu}{2})_n}{\Gamma(\frac{1-\nu}{2})(\frac{1}{2})_n}
         \Psi\left(-\frac{\nu}{2}+k, \frac{1}{2}+k; \frac{z^2}{2}\right) \right. \notag \\
         &\hspace{6.5cm}+ \frac{2^{k-1/2} \Gamma(-\frac{1}{2})(\frac{1-\nu}{2})_n}
         {z^{2k-1} \Gamma(\frac{-\nu}{2})(\frac{3}{2})_n}
         \Psi\left(-\frac{\nu}{2}, \frac{1}{2}-k; \frac{z^2}{2}\right) \bigg] \notag \\ 
         &=w_0+2^{\nu-1/2}\sum_{n=0}^{+\infty}
         \left[ \frac{\Gamma(\frac{1}{2})(-\frac{\nu}{2})_n}{\Gamma(\frac{1-\nu}{2})(\frac{1}{2})_n}
         G_{2,3}^{2,1}\left(\begin{array}{c}
         1, 1+n-\frac{\nu}{2}, - \\
         \frac{1}{2}+n, 1+n, 0 
    \end{array} \; \middle| \; \frac{z^2}{2}\right) \right. \notag \\
         &\hspace{3cm}+ \frac{\Gamma(-\frac{1}{2})(\frac{1-\nu}{2})_n}
         {\Gamma(\frac{-\nu}{2})(\frac{3}{2})_n}
         G_{2,3}^{2,1}\left(\begin{array}{c}
         1, \frac{3}{2}+n-\frac{\nu}{2}, - \\
         1+n, \frac{3}{2}+n, 0 
    \end{array} \; \middle| \; \frac{z^2}{2}\right) \bigg],
\end{align}
where $(z)_n=\frac{\Gamma(z+n)}{\Gamma(z)}$ denotes the Pochhammer symbol and $G$ is the Meijer's $G$-function. The computational implementation of $w$ through this expression is quite complicated, due to the need of considering a large number of terms in the series to achieve a good precision, which results in a considerable computation time; thus, it is worth to take into account as an alternative the differential method.

For doing that, we realize first that the parametric derivative of $D_\nu$ is given by
\begin{align}
    \frac{\partial D_\nu(z)}{\partial \nu}
    =\sqrt{\pi}e^{-\frac{z^2}{4}}
    &\left[
    \frac{2^\frac{\nu-1}{2}(\log(2)-2\psi(-\nu))z}{\Gamma(-\frac{\nu}{2})}{}_1F_1\left(\frac{1-\nu}{2};\frac{3}{2};\frac{z^2}{2}\right)
    -\frac{2^{\frac{\nu}{2}-1}(\log(2)-2\psi(-\nu))}{\Gamma(\frac{1-\nu}{2})}{}_1F_1\left(\frac{-\nu}{2};\frac{1}{2};\frac{z^2}{2}\right)\right.\notag\\
    &-\frac{2^{\frac{\nu}{2}-1}}{\Gamma(\frac{1-\nu}{2})}
    \sum_{k=0}^{+\infty}\frac{(-\frac{\nu}{2})_k\psi(k-\frac{\nu}{2})z^{2k}}
    {k!(\frac{1}{2})_k2^k}+\frac{2^{\frac{\nu-1}{2}}ze^{-\frac{z^2}{4}}}{\Gamma(-\frac{\nu}{2})}\sum_{k=0}^{+\infty}\frac{(\frac{1-\nu}{2})_k\psi(k+\frac{1-\nu}{2})z^{2k}}
    {k!(\frac{3}{2})_k2^k}\bigg],
\end{align}
with $\psi(z)$ being the digamma function~\cite{Prudnikov1989_Vol1}. Additionally, using the relationship between the parabolic cylinder function and its derivative,
\begin{equation}
    \partial_z D_\nu(z)=\frac{1}{2}zD_\nu(z)-D_{\nu+1}(z),
\end{equation}
it follows that
\begin{equation}
    W[D_\nu,\partial_\nu D_\nu](\nu, z)
    =W_\nu[D_{\nu+1},D_\nu](\nu, z)
    =D_{\nu+1}(z)\partial_\nu D_\nu(z)-\partial_\nu D_{\nu+1}(z)D_\nu(z),
\end{equation}
where $W_\nu$ represents now the Wronskian in the parameter $\nu$. Taking this into account, and the expression \cref{eq:w_differential} for $w$, we arrive at
\begin{equation}
    w(\epsilon; x)=W_0+\sqrt{\omega}\;W_\nu[D_{\nu+1},D_\nu](\nu, z).
\end{equation}
Consequently, the function $\eta$ and its partial derivatives with respect to $x$ take the following form
\begin{equation}
    \eta(\epsilon, x)
    =-\partial_x\log{w(\epsilon, x)}
    =\frac{D_\nu^2(z)}{W_0+\sqrt{\omega}W[D_\nu,\partial_\nu D_\nu](\nu, z)}\equiv\tilde\eta(\nu, z),
\end{equation}
\begin{equation}
    \partial_x \eta(\epsilon, x)
    =\sqrt{\omega}\;\tilde\eta(\nu, z)
    \left(
    z-2\frac{D_{\nu+1}(z)}{D_\nu(z)}
    +\sqrt{\omega}\;\tilde\eta(\nu, z)
    \right),
\end{equation}
\begin{align}
    \partial_x^2\eta(\epsilon, x)
    =\omega\,\tilde\eta(\nu, z)&\bigg[1+z^2+2\frac{D_{\nu+2}(z)-2zD_{\nu+1}(z)}{D_\nu(z)}+2\frac{D_{\nu+1}^2(z)}{D_\nu^2(z)}\\
    &+3\sqrt{\omega}\left(z-2\frac{D_{\nu+1}(z)}{D_\nu(z)}\right)\tilde\eta(\nu, z)+2\omega\tilde\eta^2(\nu, z)
    \bigg].
\end{align}
Summarizing, the intertwining operator $L_2^-$ used to implement the transformation, as well as the SUSY partner potential $V_2$, are determined by Eqs.~\cref{eq:intertwining_operator} and \cref{eq:SUSY_partner_potential}, respectively. Meanwhile, the eigenfunctions $\psi_n^{(2)}$ are derived from expression \cref{eq:eigenfunctions_H2}, with the exception of the missing state which is given by Eq.~\cref{eq:missing_state}.

We conclude that, by choosing a factorization energy within the spectrum of $H_0$, $\epsilon = \mathcal{E}_j$, in the regular case the confluent SUSY transformation is isospectral. Nevertheless, in the critical case it turns out that $\text{Sp}(H_2) = \text{Sp}(H_0) \setminus \{\mathcal{E}_j\}$, i.e., the transformation \textit{deletes} the level $\mathcal{E}_j$ from the spectrum of $H_2$. On the other hand, for $\epsilon \neq \mathcal{E}_j$ in the regular case $\text{Sp}(H_2) = \text{Sp}(H_0) \cup \{\epsilon\}$, i.e., the transformation \textit{creates} a new energy level for $H_2$ at $\epsilon$. However, in the critical case the transformation becomes isospectral.

\section{Confluent algorithm applied to bilayer graphene}
\label{sec:confluent_algorithm_bilayer_graphene}

Let us take as initial potential the shifted harmonic oscillator $V_0$ of Eq.~\cref{eq:harmonic_oscillator_potential}, and let us express the factorization energy used to implement the second-order confluent algorithm as $\epsilon = \omega \nu$, $\nu \in \mathbb{R}$. Taking into account the spectrum of $H_0$ given by Eq.~\cref{eq:solutions_HO}, we can deduce from Eq.~\cref{eq:eigenvalues_BG} that the eigenenergies for the electrons in bilayer graphene (BG) are given by 
\begin{equation}
\label{eq:spectrum_bilayer_graphene}
    E_n=\hbar\omega_0\,\abs{n-\nu},
    \hspace{3mm} n = 0, 1, 2, \ldots, \quad \text{with} \quad 
    \omega_0:=\frac{\hbar\omega}{2m^*},
\end{equation}
where the inclusion of $E_\nu=0$ in the bilayer graphene spectrum $\text{Sp}(\mathbb{H})$ depends on the factorization energy chosen and the square integrability of the missing state \cref{eq:missing_state}. According to the discussion of \cref{subsec:confluent_algorithm_harmonic_oscillator} for $\epsilon = \mathcal{E}_j$ we have that $E_\nu = E_j=0$ constitutes an energy level of bilayer graphene in both regular and critical cases. On the other hand, for $\epsilon \neq \mathcal{E}_j$ ($\nu\neq j$) it turns out that $E_\nu = 0 \in \text{Sp}(\mathbb{H})$ in the regular case, whereas $E_\nu = 0 \notin \text{Sp}(\mathbb{H})$ in the critical case. Taking this into account, we get the four different cases reported in \cref{tab:cases}.\\ 
\begin{table}[hbt]
\centering
\renewcommand{\arraystretch}{1.3}
\begin{tabular}{ccccc}
\toprule
\textbf{Condition on $\epsilon$} & & \textbf{$w_0$ Value} & & \textbf{Case} \\ 
\midrule
\multirow{2}{*}{$\epsilon = \mathcal{E}_j$} 
    & & $w_r$ & & \caseCI \label{case:CI} \\ 
    & & $w_c$ & & \caseCII \label{case:CII} \\ 
\midrule
\multirow{2}{*}{$\epsilon \neq \mathcal{E}_j$}    
    & & $w_r$ & & \caseCIII \label{case:CIII} \\ 
    & & $w_c$ & & \caseCIV \label{case:CIV} \\ 
\bottomrule
\end{tabular}
\caption{Confluent algorithm classification based on the values of $\epsilon$ and $w_0$.}
\label{tab:cases}
\end{table}\\
For the four general cases of \cref{tab:cases}, the eigenstates of bilayer graphene associated to the eigenvalues \cref{eq:spectrum_bilayer_graphene} are given by
\begin{equation}
\label{eq:eigenfunctions_BG}
    \Psi_n(x,y) = \frac{e^{iky}}{\sqrt{2}}
\begin{pmatrix}
    \displaystyle
    \frac{L_2^-\psi_n^{(0)}(x)}{\abs{\mathcal{E}_n-\epsilon}}\\[1em]
    \psi_n^{(0)}(x)
\end{pmatrix}, \quad n\in\mathbb{N}_0\backslash\{\nu\}.
\end{equation}
Additionally, the eigenstate associated with $E_\nu = 0$ (which does not exist in case CIV) is
\begin{equation}
    \Psi_\nu(x,y)=e^{iky}\left\{
\begin{array}{ll}
\frac{1}{\sqrt{2}}
\begin{pmatrix}
\psi_{\epsilon}^{(2)}(x) \\[0.5em]
\psi_j^{(0)}(x)
\end{pmatrix} & \text{\caseCI}\\[2em]
\begin{pmatrix}
0 \\[0.5em]
\psi_j^{(0)}(x)
\end{pmatrix} & \text{\caseCII}\\[2em] 
\begin{pmatrix}
\psi_{\epsilon}^{(2)}(x) \\[0.5em] 
0
\end{pmatrix} & \text{\caseCIII}
\end{array}
\right..
\end{equation}
Furthermore, the completeness relation for the eigenstates of bilayer graphene takes the form
\begin{equation}
\label{eq:CR_III}
    \ket{\Psi_\nu}\bra{\Psi_\nu}
    +\sum_{n=0}^{+\infty}\ket{\Psi_n}\bra{\Psi_n}=\mathbb{1},
\end{equation}
where $\mathbb{1}$ is the identity operator in the state space of bilayer graphene, and the first term $\ket{\Psi_\nu}\bra{\Psi_\nu}$ in the left-hand side must be neglected in cases \caseCI, \caseCII and \caseCIV.

\subsection{Energy levels}
\label{subsec:energy_levels}

Note that most properties of the bilayer graphene spectrum, such as ordering, spacing and degeneracy, depend on the factorization energy $\epsilon$ chosen, as well as on the interval to which the parameter $w_0$ belongs.

\subsubsection{Ordering}
\label{subsubsec:ordering}
In general, the natural ordering of the energy levels of bilayer graphene does not correspond with the labeling in terms of $n$. In fact, for $\epsilon < \omega/2$, the ordering induced by $n$ is standard: in the regular case, the energy levels are given by
\begin{equation}
    \big\{
    E_\nu=0,
    E_0=\abs{\nu}\,\hbar\omega_0,\;
    E_1=\abs{1-\nu}\,\hbar\omega_0,\;
    E_2=\abs{2-\nu}\,\hbar\omega_0,\;\ldots\big\},
\end{equation}
while in the critical case they become
\begin{equation}
    \big\{
    E_0=\abs{\nu}\,\hbar\omega_0,\;
    E_1=\abs{1-\nu}\,\hbar\omega_0,\;
    E_2=\abs{2-\nu}\,\hbar\omega_0,\;\ldots\big\}.
\end{equation}
On the other hand, for $\epsilon \geq \omega/2$ such natural ordering is lost, depending on the interval $(\mathcal{E}_j, \mathcal{E}_{j+1})$ on which $\epsilon$ lies. For instance, for $\epsilon = 0.7\,\omega$ the energies in the regular case are
\begin{equation}
    \big\{
    E_\nu=0,\;
    E_0=0.7\,\hbar\omega_0,\;
    E_1=0.3\,\hbar\omega_0,\;
    E_2=1.3\,\hbar\omega_0,\;
    E_3=2.3\,\hbar\omega_0,\;\ldots\big\}.
\end{equation}
Meanwhile, if $\epsilon=2\omega$ in both regular and critical cases they become
\begin{equation}
    \big\{
    E_0=E_4=2\,\hbar\omega_0,\;
    E_1=E_3=1\,\hbar\omega_0,\;
    E_2=0\;
    E_5=3\,\hbar\omega_0,\;\ldots\big\}.
\end{equation}
Although in the last two cases the natural ordering is lost, we can always define an alternative label $N$ that reflects the correct ordering. \cref{fig:energy_levels} shows the energy levels of bilayer graphene as function of the index of the $H_0$-eigenvalues for $\epsilon=0.7\,\omega$ and $\epsilon=2\,\omega$.

\subsubsection{Spacing}
\label{subsubsec:spacing}

From Eq.~\cref{eq:spectrum_bilayer_graphene} we can deduce that for $\epsilon = \mathcal{E}_j$ the bilayer graphene spectrum becomes equidistant in both the regular and critical cases, as shown in \cref{fig:levels_ϵ=2}. On the other hand, for $0 < \epsilon \neq \mathcal{E}_j$, the level spacing is uniform only in the critical case with $\epsilon = \mathcal{E}_j - \omega/2$, and is not uniform otherwise. Now, for $\epsilon < 0$ the energy levels are equidistant in the critical case, while in the regular case they are equidistant only when $\epsilon = -\mathcal{E}_1$.

However, in the cases where the spacing is not entirely uniform the total state space of bilayer graphene consists of two subspaces, one finite-dimensional and another infinite-dimensional. The finite-dimensional subspace may contain one or more equidistant energy levels, while the level spacing in the infinite-dimensional subspace is certainly equidistant. In these cases we say that the bilayer graphene spectrum is partially equidistant.

\subsubsection{Degeneracy}
\label{subsubsec:degeneracy_levels}

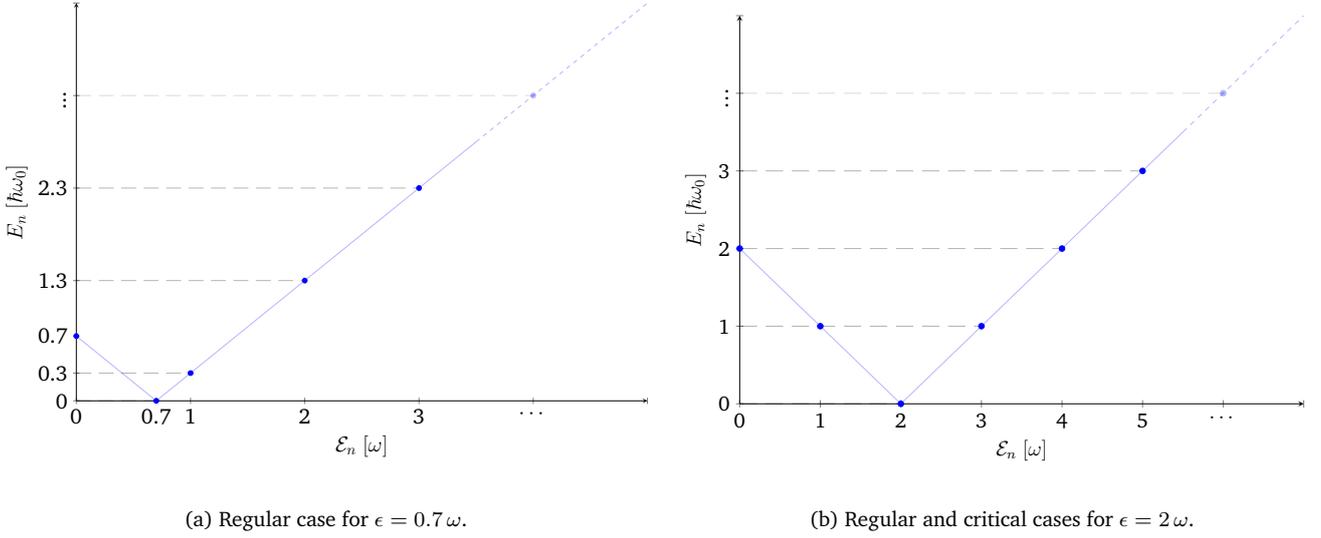
\begin{figure}[bt]
    \centering
    \begin{subfigure}[b]{0.49\textwidth}
        \centering
        \begin{tikzpicture}[scale=0.56]
\begin{axis}[
    xlabel={$\mathcal{E}_n$ $[\omega]$},
    ylabel={$E_n$ $[\hbar\omega_0]$},
    xtick={0, 0.7, 1, 2, 3, 4, 5},
    xticklabels={0, 0.7, 1, 2, 3, $\dotsb$},
    ytick={0, 0.3, 0.7, 1.3, 2.3, 3.3, 4.3},
    yticklabels={0, 0.3, 0.7, 1.3, 2.3, \vdots },
    ymin=0, ymax=4.3,
    xmin=0, xmax=5,
    width=15cm, height=11cm,
    axis lines=left,
    xlabel style={font=\Large}, 
    ylabel style={font=\Large}, 
    ticklabel style={font=\Large},
]

\addplot[
    only marks, 
    mark=*,
    mark options={scale=0.8, fill=blue, draw=blue}, 
] coordinates {
        (0, 0.7)
        (0.7, 0)
        (1, 0.3)
        (2, 1.3)
        (3, 2.3)
};
\addplot[
    only marks, 
    mark=*,
    mark options={opacity=0.3, scale=0.8, fill=blue, draw=blue}, 
] coordinates { (4, 3.3)};

\addplot[color=blue, opacity=0.3] coordinates{(0, 0.7) (0.7, 0)};
\addplot[color=blue, opacity=0.3] coordinates{(0.7, 0) (3.5, 2.8)};
\addplot[color=blue, opacity=0.3, dashed] coordinates{(3.5, 2.8) (5,4.3)};

\draw[dashed, semithick, opacity=0.3, dash pattern=on 10pt off 5pt]
(axis cs:0, 0.7) -- (axis cs:0, 0.7);
\draw[dashed, semithick, opacity=0.3, dash pattern=on 10pt off 5pt]
(axis cs:0, 0) -- (axis cs:0.7, 0);
\draw[dashed, semithick, opacity=0.3, dash pattern=on 10pt off 5pt]
(axis cs:0, 0.3) -- (axis cs:1, 0.3);
\draw[dashed, semithick, opacity=0.3, dash pattern=on 10pt off 5pt]
(axis cs:0, 1.3) -- (axis cs:2, 1.3);
\draw[dashed, semithick, opacity=0.3, dash pattern=on 10pt off 5pt]
(axis cs:0, 2.3) -- (axis cs:3, 2.3);
\draw[dashed, semithick, opacity=0.15, dash pattern=on 10pt off 5pt]
(axis cs:0, 3.3) -- (axis cs:4, 3.3);

\end{axis}
\end{tikzpicture}
        \caption{Regular case for $\epsilon=0.7\,\omega$.}
        \label{fig:levels_ϵ=0.7}
    \end{subfigure}
    \hfill
    \begin{subfigure}[b]{0.49\textwidth}
        \centering
        \begin{tikzpicture}[scale=0.65]
\begin{axis}[
    xlabel={$\mathcal{E}_n$ $[\omega]$},
    ylabel={$E_n$  $[\hbar\omega_0]$},
    xtick={0, 1, ..., 7},
    xticklabels={0, 1, 2, 3, 4, 5, $\dotsb$},
    ytick={0, ..., 7},
    yticklabels={0, 1, 2, 3, \vdots },
    ymin=0, ymax=5,
    xmin=0, xmax=7,
    width=13cm, height=9.5cm,
    axis lines=left,
    xlabel style={font=\large}, 
    ylabel style={font=\large}, 
    ticklabel style={font=\large},
]

\addplot[
    only marks, 
    mark=*,
    mark options={scale=0.8, fill=blue, draw=blue}, 
] coordinates {
        (0, 2)
        (1, 1)
        (2, 0)
        (3, 1)
        (4, 2)
        (5, 3)
};
\addplot[
    only marks, 
    mark=*,
    mark options={opacity=0.3,scale=0.8, fill=blue, draw=blue}, 
] coordinates {
        (6, 4)
};

\addplot[color=blue, opacity=0.3] coordinates{(0, 2) (2, 0)};
\addplot[color=blue, opacity=0.3] coordinates{(2, 0) (5.5, 3.5)};
\addplot[color=blue, opacity=0.3, dashed] coordinates{(5.5, 3.5) (7, 5)};

\draw[dashed, semithick, opacity=0.15, dash pattern=on 10pt off 5pt]
(axis cs:0, 4) -- (axis cs:6, 4);
\draw[dashed, semithick, opacity=0.3, dash pattern=on 10pt off 5pt]
(axis cs:0, 3) -- (axis cs:5, 3);
\draw[dashed, semithick, opacity=0.3, dash pattern=on 10pt off 5pt]
(axis cs:0, 2) -- (axis cs:4, 2);
\draw[dashed, semithick, opacity=0.3, dash pattern=on 10pt off 5pt]
(axis cs:0, 1) -- (axis cs:3, 1);
\draw[dashed, semithick, opacity=0.3, dash pattern=on 10pt off 5pt]
(axis cs:0, 0) -- (axis cs:2, 0);

\end{axis}
\end{tikzpicture}
        \vspace{10pt}
        \caption{Regular and critical cases for $\epsilon=2\,\omega$.}
        \label{fig:levels_ϵ=2}
    \end{subfigure}
    \caption{Energy levels of bilayer graphene as a function of the level index of $H_0$. The natural ordering of the BG levels does not correspond with the standard ordering of the auxiliary problems.}
    \label{fig:energy_levels}
\end{figure}

In case that degeneracy appears, for a factorization energy that is a positive integer or half-integer multiple of $\omega$, there are a finite number of degenerate energy levels. In fact, if either $\epsilon = \mathcal{E}_j$ or $\epsilon = \mathcal{E}_j - \omega/2$ with $j > 0$, the first $j$ excited states are two-fold degenerate, as illustrated in \cref{fig:levels_ϵ=2} for $\epsilon = 2\,\omega$. Moreover, $E_\nu = 0$ turns out to be a non-degenerate ground state energy level, except in the critical case with $\epsilon \neq \mathcal{E}_j$ for which $E_\nu=0$ does not belong to the bilayer graphene spectrum. In particular, in the critical case for $\epsilon = \mathcal{E}_j - \omega/2$ ($j > 0$) the ground state level is two-fold degenerate, whereas in any other case it is non-degenerate.

Therefore, the bilayer graphene spectrum is non-degenerate and equidistant with standard ordering in the regular case when $\epsilon = \mathcal{E}_0$ or $\epsilon = -\mathcal{E}_1$, and in the critical case when $\epsilon \leq 0$. 

\subsection{Probability density and probability current}
\label{subsec:probability_current_bilayer_graphene}
The probability density for an arbitrary state of bilayer graphene is given by
\begin{equation}
\label{eq:density_BG}
    \rho
    =\Psi^\dagger\Psi,
\end{equation}
while the probability current, as reported in \cite{Fernandez2021_BGMagneticFields}, can be written as follows:
\begin{equation}
\label{eq:xcurrent_BG}
    J_x
    =\dfrac{\hbar}{m^*}\text{Im}\left[
    \Psi^\dagger
    \left(
    \sigma_x\partial_x+\sigma_y\partial_y
    +\dfrac{ie}{\hbar c}A(x)\sigma_y
    \right)\Psi
    \right],
\end{equation}
\begin{equation}
\label{eq:ycurrent_BG}
    J_y
    =\dfrac{\hbar}{m^*}\text{Im}\left[
    \Psi^\dagger
    \left(
    \sigma_y\partial_x-\sigma_x\partial_y
    -\dfrac{ie}{\hbar c}A(x)\sigma_x
    \right)\Psi
    \right],
\end{equation}
where $\sigma_j$, $j=x,y,z$ are the Pauli matrices.
In particular, for the eigenstates \cref{eq:eigenfunctions_BG}, it turns out that
\begin{align}
    \rho_n(x) 
    &= \dfrac{1}{2}\left(
    \abs{\psi_n^{(2)}(x)}^2 + \abs{\psi_n^{(0)}(x)}^2
    \right), \\[1em]
    J_{x,n}(x) &= 0, \\[0.5em]
    J_{y,n}(x)
    &=\dfrac{\hbar}{2m^*}\left[
    W[\psi_n^{(0)},\psi_n^{(2)}](x)-\eta(x)\psi_n^{(0)}(x)\psi_n^{(2)}(x)
    \right],
\end{align}
for any $n\in\mathbb{N}_0\cup\{\nu\}$.

\begin{figure}[p]
    \centering
    \begin{subfigure}[b]{0.49\textwidth}
        \centering
        \includegraphics[width=\textwidth]{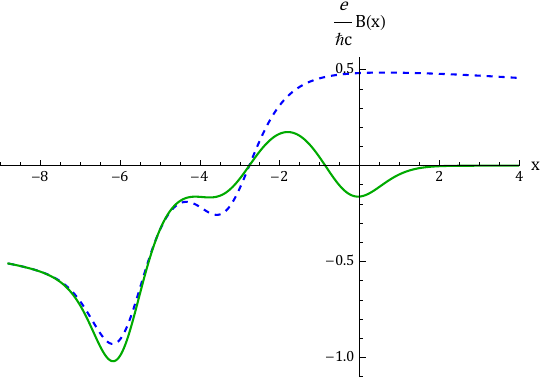}
        \caption{$\epsilon=\omega/2$}
        \label{fig:magnetic_field_1/2}
    \end{subfigure}
    \hfill
    \begin{subfigure}[b]{0.49\textwidth}
        \centering
        \includegraphics[width=\textwidth]{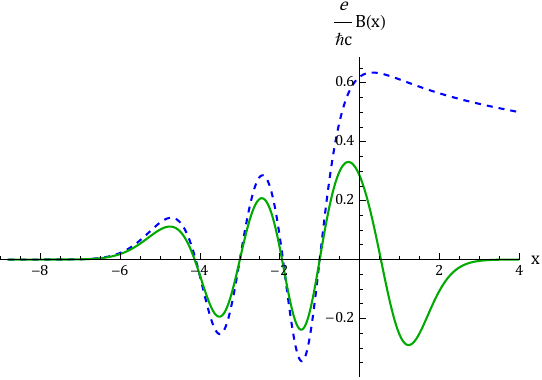}
        \caption{$\epsilon=2\omega$}
        \label{fig:magnetic_field_2}
    \end{subfigure}\caption{Applied magnetic field for two different factorization energies: $\epsilon=\omega/2$ (left) and $\epsilon=2\omega$ (right). It is shown the critical case for $w_0=0$ (dotted blue lines) and the regular case for $w_0=1$ (green lines).}
    \label{fig:magnetic_fields}
\end{figure}

\begin{figure}[p]
    \centering
    \begin{subfigure}[b]{0.49\textwidth}
        \centering
        \includegraphics[width=\textwidth]{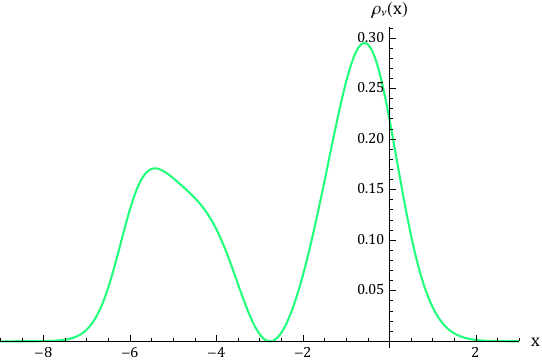}
        \caption{Probability density of the non-degenerate ground state eigenfunction for $w_0=1$ (regular case).}
        \label{fig:density_w=1}
    \end{subfigure}
    \hfill
    \begin{subfigure}[b]{0.49\textwidth}
        \centering
        \includegraphics[width=\textwidth]{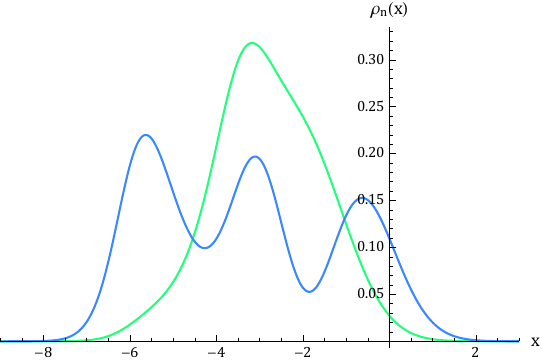}
        \caption{Ground state probability density in the critical case ($w_0=0$). Two orthogonal eigenfunctions $\Psi_0$ (green) and $\Psi_1$ (blue) are associated with the ground state level.}
        \label{fig:density_w=0}
    \end{subfigure}
    \caption{Probability density of the bilayer graphene ground state eigenfunction for $\epsilon = \omega/2$ in the regular (a) and critical case (b).}
    \label{fig:densities_ϵ=0.5}
\end{figure}

\begin{figure}[p]
    \centering
    \begin{subfigure}[b]{0.49\textwidth}
        \centering
        \includegraphics[width=\textwidth]{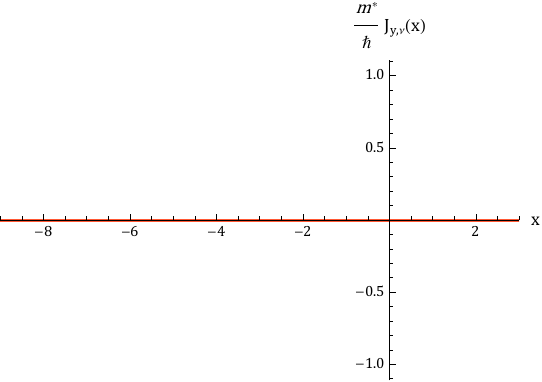}
        \caption{Regular case with $w_0=1$: the probability current in $y$-direction for the ground state vanishes.}
        \label{fig:currentY_w=1}
    \end{subfigure}
    \hfill
    \begin{subfigure}[b]{0.49\textwidth}
        \centering
        \includegraphics[width=\textwidth]{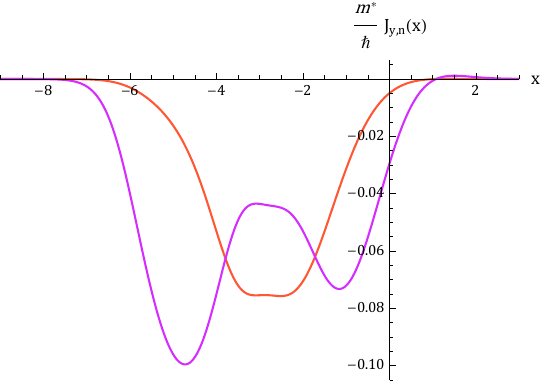}
        \caption{Critical case with $w_0=0$: probability current for $\Psi_0$ (orange) and $\Psi_1$ (purple).}
        \label{fig:currentY_w=0}
    \end{subfigure}
    \caption{Probability current in $y$-direction for the ground state with $\epsilon = \omega/2$.}
    \label{fig:currentsY_ϵ=0.5}
\end{figure}

Let us consider next an interesting particular case to exemplify the previous results.
\newpage
\subsection{Case with \texorpdfstring{$\epsilon=\omega/2$}{epsilon 1/2}}

Let us choose $\epsilon = \omega/2$ as factorization energy.\footnote{In the remainder of this paper, when we illustrate the graphs for specific examples we will set $\omega=1$ and $k=1$.} Each $w_0 \in [0, +\infty)$ defines unambiguously a non-singular external magnetic field $B$, as well as the spectrum and the set of eigenstates for the corresponding bilayer graphene Hamiltonian. For instance, in the regular case with $w_0 = 1$ the electron energies for bilayer graphene turn out to be
\begin{equation}
\label{eq:eigenvalues_BG_ϵ=0.5}
    E_n = \hbar\omega_0 |n - 1/2|,
    \quad n = 1/2, 0, 1, 2, \ldots\;, 
\end{equation}
where $E_{1/2} = 0$ is the ground state energy. In this case the bilayer graphene spectrum is partially equidistant in the full state space, but in the subspace spanned by $\{\ket{\Psi_n}\}_{n=0}^{\infty}$ the energy levels are equidistant. Moreover, the first excited state $E_0 = E_1 = \hbar\omega_0/2$ is two-fold degenerate, with two orthogonal eigenfunctions $\Psi_0$ and $\Psi_1$; meanwhile, the remaining levels are non-degenerate. 

In contrast,  in the critical case with $w_0 = 0$ the energy level $E_{1/2}=0$ is no longer in the spectrum of the bilayer graphene Hamiltonian, thus all energy levels are equidistant. Additionally, the first excited state of the regular case becomes the ground state in the critical case, which now is the only two-fold degenerate level.

The external magnetic field in the two preceding cases is illustrated in \cref{fig:magnetic_field_1/2}. The probability density and probability current in $y$-direction for both cases are shown in \cref{fig:densities_ϵ=0.5,fig:currentsY_ϵ=0.5}, respectively. Note that the current density in $x$-direction vanishes for any eigenstate of the bilayer graphene Hamiltonian.

\section{Bilayer graphene coherent states}
\label{sec:coherent_states_bilayer_graphene}

In this section, we will derive new coherent states for bilayer graphene in position-dependent magnetic fields generated through the confluent second-order SUSY QM. Unlike the previously constructed CS in the real case for factorization energies  $\epsilon_1=\mathcal{E}_0,\epsilon_2=\mathcal{E}_1\in\text{Sp}(H_0)$ for bilayer graphene placed a in constant homogeneous magnetic field \cite{Fernandez2020_BG_CoherentStates,Fernandez2022_GGCoherentStates}, these new coherent states are derived for any real factorization energy, $\epsilon \in \mathbb{R}$. Furthermore, we will show the existence of factorization energies for which these states exhibit exact temporal stability.

\subsection{Ladder operators for bilayer graphene}
\label{subsec:ladder_operators_bilayer_graphene}

The ladder operators for bilayer graphene can be constructed from those of the two auxiliary Schrödinger problems \cite{DiazBautista2017_GraphenCoherentStates,Fernandez2020_BG_CoherentStates}, i.e., from the ladder operators of the shifted harmonic oscillator of \cref{subsec:confluent_algorithm_harmonic_oscillator},
\begin{equation}
\label{eq:LO_HO}
    a^{\pm} = \dfrac{1}{\sqrt{\omega}}\left(\mp\dfrac{d}{dx} + \dfrac{\omega}{2}x + k\right),
\end{equation}
which satisfy the Heisenberg-Weyl algebra, and from deformations of the natural ladder operators of the SUSY partner potential characterized by an arbitrary function $f$ of the form $B^- = L_2^- a^- f(H_0) L_2^+$ and $B^+ \equiv (B^-)^\dagger$, which fulfill a polynomial Heisenberg algebra \cite{Fernandez2004_HigherOrder,Fernandez2020_BG_CoherentStates}.

The simplest ladder operators for bilayer graphene are diagonal $2\times2$ matrix operators, such that each component of the resultant state depends just on the corresponding initial state component \cite{Fernandez2022_GGCoherentStates}. Thus, we introduce the following diagonal lowering operator:
\begin{equation}
\label{eq:lowering_operator_diag}
    \mathbb{A}^-=
    \begin{pmatrix}
        L_2^-\dfrac{1}{\abs{H_0-\epsilon}}
        a^- \dfrac{f(H_0)}{\abs{H_0-\epsilon}}L_2^+ & 0\\[1.5em]
        0 & a^- f(H_0)
    \end{pmatrix},
\end{equation}
where $f$ is an arbitrary real function of $H_0$. Meanwhile, the raising operator is $\mathbb{A}^+:=\big(\mathbb{A}^-\big)^\dagger$. It is worth to notice that there may exist eigenstates that cannot be connected through such diagonal ladder operators, but they can be linked using non-diagonal ones, thus making worthwhile to consider as well this case. A non-diagonal version of the lowering operator is given by
\begin{equation}
\label{eq:lowering_operator_non_diag}
    \mathbb{\widetilde A}^-=\dfrac{1}{2}
    \begin{pmatrix}
        L_2^-\dfrac{1}{\abs{H_0-\epsilon}}a^-\dfrac{f(H_0)}{\abs{H_0-\epsilon}}L_2^+ & L_2^-\dfrac{1}{\abs{H_0-\epsilon}}a^-f(H_0)\\[1.5em]
        a^- \dfrac{f(H_0)}{\abs{H_0-\epsilon}}L_2^+ & a^- f(H_0)
    \end{pmatrix},
\end{equation}
while the non-diagonal raising operator is the adjoint of $\mathbb{\widetilde A}^-$.

\subsubsection{Action of the ladder operators for \texorpdfstring{$\epsilon = \mathcal{E}_j$}{epsilon equal to Ej}}
\label{subsec:action_CI}

The action of the previous ladder operators onto the eigenstates \cref{eq:eigenfunctions_BG} in the first two cases of \cref{tab:cases} becomes
\begin{align}
    \label{eq:action_ladder_operators_J_+}
    \mathbb{A}^-\ket{\Psi_n}
    &= f(\mathcal{E}_n)\sqrt{n}\ket{\Psi_{n-1}},
    \quad \forall\,n\in\mathbb{N}_0\backslash\mathcal{J}_+,\\[0.5em]
    \label{eq:action_ladder_operators_J_-}
    \mathbb{A}^+\ket{\Psi_n}
    &= f(\mathcal{E}_{n+1})\sqrt{n+1}\ket{\Psi_{n+1}},
    \quad \forall\,n\in\mathbb{N}_0\backslash\mathcal{J}_-,
\end{align}
where $\mathcal{J}_\pm=\{j,j\pm1\}$. Besides, since the seed solution $\psi_j^{(0)}$ lies in the kernel of $L_2^-$, it follows that
\begin{align}
    \label{eq:restriction_II}
    \mathbb{A}^-\ket{\Psi_{j+1}} &\not\propto \ket{\Psi_j},  & \mathbb{A}^+\ket{\Psi_{j-1}} &\not\propto \ket{\Psi_j}, & \text{\caseCI} \\[0.5em]
    \label{eq:action_CII_n=j+1}
    \mathbb{A}^-\ket{\Psi_{j+1}} &= \dfrac{f(\mathcal{E}_{j+1})}{\sqrt{2}}\sqrt{j+1}\ket{\Psi_j}, & \mathbb{A}^+\ket{\Psi_{j-1}} &= \dfrac{f(\mathcal{E}_j)}{\sqrt{2}}\sqrt{j}\ket{\Psi_j}. & \text{\caseCII}
\end{align}
Let us notice that Eqs.~\cref{eq:action_ladder_operators_J_+,eq:action_ladder_operators_J_-,eq:restriction_II,eq:action_CII_n=j+1} hold as well in the non-diagonal case. Moreover, since the missing state $\psi_j^{(2)}$ is annihilated by $L_2^+$ in case \caseCI, and the upper component of $\ket{\Psi_j}$ is zero in \caseCII, it turns out that 
\begin{equation}
\label{eq:restriction_I}
    \mathbb{A}^-\ket{\Psi_j} \not\propto \ket{\Psi_{j-1}},
    \qquad \mathbb{A}^+\ket{\Psi_j} \not\propto \ket{\Psi_{j+1}},
\end{equation}
in both \caseCI and \caseCII cases. However, due to the off-diagonal components of $\mathbb{\widetilde A}^\mp$ we have
\begin{equation}
    \mathbb{\widetilde A}^-\ket{\Psi_j} = \alpha f(\mathcal{E}_j)\sqrt{j}\ket{\Psi_{j-1}}, \qquad 
    \mathbb{\widetilde A}^+\ket{\Psi_j} = \alpha f(\mathcal{E}_{j+1})\sqrt{j+1}\ket{\Psi_{j+1}},
\end{equation}
where $\alpha = 1/2$ for \caseCI and $\alpha = 1/\sqrt{2}$ for \caseCII. Therefore, in the critical case the non-diagonal ladder operators manage to connect all the eigenstates of the bilayer graphene Hamiltonian. Meanwhile, in the other cases the action of the ladder operators carries some eigenstates outside the state space. In such cases, in order for the ladder operators to leave invariant the state space of bilayer graphene, we must choose $f(\mathcal{E}_j) = f(\mathcal{E}_{j+1}) = 0$.

\subsubsection{Action of the ladder operators for \texorpdfstring{$\epsilon \neq \mathcal{E}_j$}{epsilon not equal to Ej}}
\label{subsubsec:action_CIII}

In the last two cases of \cref{tab:cases} the action of the ladder operators of bilayer graphene (both diagonal and non-diagonal) turns out to be
\begin{align}
\label{eq:action_A^-_diag}
    \mathbb{A}^-\ket{\Psi_n}
    &= f(\mathcal{E}_n)\sqrt{n}\ket{\Psi_{n-1}},
    \quad \forall\,n\in\mathbb{N}_0,
    &\mathbb{A}^-\ket{\Psi_\nu}=0,\\[0.5em]
    \label{eq:action_A^+_diag}
    \mathbb{A}^+\ket{\Psi_n}
    &= f(\mathcal{E}_{n+1})\sqrt{n+1}\ket{\Psi_{n+1}},
    \quad \forall\,n\in\mathbb{N}_0,
    &\mathbb{A}^+\ket{\Psi_\nu}=0.
\end{align}
Note that in case \caseCIV the expressions to the right must be disregarded, since $\ket{\Psi_\nu}$ is not a bilayer graphene eigenstate in that case. Moreover, in case \caseCIII $\ket{\Psi_\nu}$ is an \textit{isolated} eigenstate, since it is annihilated by both $\mathbb{A}^\mp$. Nevertheless, this will not affect the coherent states derivation since all other eigenstates are connected to each other through the ladder operators.

\subsubsection{General action of the ladder operators}
\label{subsec:action_general_form}

In general, the action of the ladder operators in all previous cases can be written as follows
\begin{align}
\label{eq:action_general_1}
    \mathbb{A}^-\ket{\Psi_n}
    &= g(\mathcal{E}_n)\sqrt{n}\ket{\Psi_{n-1}},
    \quad \forall\,n\in\mathbb{N}_0,\\[0.5em]
    \label{eq:action_general_2}
    \mathbb{A}^+\ket{\Psi_n}
    &= g(\mathcal{E}_{n+1})\sqrt{n+1}\ket{\Psi_{n+1}},
    \quad \forall\,n\in\mathbb{N}_0,
\end{align}
where $g = f$, except in the non-diagonal critical case with $\epsilon = \mathcal{E}_j$ for which $g(\mathcal{E}_n) = f(\mathcal{E}_n)$ for $n \in \mathbb{N}_0 \setminus \mathcal{J}_+$ and $g(\mathcal{E}_n) = f(\mathcal{E}_n)/\sqrt{2}$ for $n = j, j+1$. Additionally, for $\epsilon = \mathcal{E}_j$ we must have $g(\mathcal{E}_j) = g(\mathcal{E}_{j+1}) = 0$ in the diagonal case (both regular and critical), as well as in the non-diagonal regular case.

For the derivation of the bilayer graphene coherent states we will consider the general case where $\{\mathcal{E}_{k_1}, \mathcal{E}_{k_2}, \ldots, \mathcal{E}_{k_l} \equiv \mathcal{E}_{k}\}$ denotes the ordered set of roots of $g$. The case without roots, with $g$ taking a finite value at $\mathcal{E}_0$, and with just one root at $\mathcal{E}_0$ will be equivalent.

\subsection{Barut-Girardello coherent states}
\label{sec:girardello_CS}

The Barut-Girardello coherent states (BGCS) are eigenstates of the lowering operator $\mathbb{A}^-$, namely, $\mathbb{A}^-\ket{\Psi_\alpha} = \alpha \ket{\Psi_\alpha}$, $\alpha \in \mathbb{C}$. They are expanded in terms of the bilayer graphene eigenstates as follows
\begin{equation}
\label{eq:expansion_I}
    \ket{\Psi_\alpha}=a_\nu\ket{\Psi_\nu} + \sum_{n=0}^{+\infty}a_n\ket{\Psi_n},
\end{equation}
where $a_\nu=0$ in cases \caseCI, \caseCII and \caseCIV. Taking into account the BGCS definition, as well as the action \cref{eq:action_general_1} of the lowering operator $\mathbb{A}^-$, the coefficients in the expansion \cref{eq:expansion_I} become
\begin{equation}
\label{ec:recurrence_relation_k}
    \begin{array}{c@{\quad}l}
         a_n = 0, & \quad n=\nu, 0,\dotsc,k-1\;,\\[0.5em]
         a_{n+k}
         = \dfrac{\alpha^{n}}{\big[g(\mathcal{E}_{n+k})\big]!\sqrt{(n+k)!/k!}}a_k,
         & \quad n \in \mathbb{N},
    \end{array}
\end{equation}
with 
\begin{equation}
    \big[g(\mathcal{E}_{n+k})\big]! := \left\{
    \begin{array}{cc}
         1 & \quad \text{for} \quad n = 0 \\[0.5em]
         g(\mathcal{E}_{k})g(\mathcal{E}_{1+k}) \dotsm g(\mathcal{E}_{n+k})
         & \quad \text{for} \quad n\in \mathbb{N}
    \end{array},
    \right.
\end{equation}
being the generalized factorial function. Thus, the BGCS turn out to be
\begin{equation}
\label{eq:coherent_states_k}
    \ket{\Psi_\alpha^{\,k}}
    =N(k,r)\sum_{n=0}^{+\infty}\frac{\alpha^n}{G(k,n)}\ket{\Psi_{n+k}},
\end{equation}
where $\alpha=re^{i\theta}$, the function $G$ and the normalization constant $N$ are given by
\begin{equation}
\label{eq:constant_BG}
    G(k,n)=\sqrt{(n+k)!}\big[g(\mathcal{E}_{n+k})\big]!\,,\qquad
    N(k,r)=\left(\sum_{n=0}^{+\infty}\frac{r^{2n}}{G^2(k,n)}\right)^{-\frac{1}{2}}.
\end{equation}

Let us stress that the fact that $g$ has a root at $\mathcal{E}_k$ implies that the first $k+1$ coefficients of the expansion \cref{eq:expansion_I} vanish, thus the BGCS can be made an overcomplete set in the subspace $\mathcal{H}_k^\infty := \text{span}\Big(\big\{\ket{\Psi_k}, \ket{\Psi_{1+k}}, \ldots, \ket{\Psi_{n+k}}, \ldots \big\}\Big)$. Furthermore, since it is not possible to generate a coherent states family in any finite-dimensional subspace through the Barut-Girardello definition (see e.g. \cite{Bermudez2014_PainleveIV}), it is more convenient to use the Gilmore-Perelomov definition in such a case.

\subsection{Gilmore-Perelomov coherent states}
\label{subsec:PCS}

Since $g$ has a finite set of roots, the commutator of each pair of ladder operators in general is not the identity. Consequently, we cannot use the Baker-Hausdorff formula to factorize the \textit{displacement} operator $\mathcal{D}(\alpha) = e^{\alpha \mathbb{A}^+ - \alpha^* \mathbb{A}^-}$. However, we can derive a kind of Gilmore-Perelomov coherent states (GPCS) by applying the non-unitary operator $D(\alpha) = e^{\alpha \mathbb{A}^+}$ or $D^+(\alpha) = e^{\alpha^* \mathbb{A}^-}$ onto an appropriate extremal state. In the following, we will consider only the first operator $D(\alpha)$; the use of the second operator $D^+(\alpha)$ is somehow equivalent.

Note that in case that two roots are consecutive, $\mathcal{E}_{k_{i+1}} - \mathcal{E}_{k_i} = \omega$, we cannot connect the corresponding eigenstates through the ladder operators. Then, the action of $D(\alpha)$ onto the \textit{lower} eigenstate $\ket{\Psi_{k_i}}$, or the action of $D^+(\alpha)$ onto the \textit{upper} one $\ket{\Psi_{k_{i+1}}}$, will produce the same state. Therefore, from now on we will consider the non-trivial case with non-consecutive roots, i.e., $\mathcal{E}_{k_{i+1}} - \mathcal{E}_{k_i} \geq 2\omega$. Thus, the GPCS obtained by applying $D(\alpha)$ to the extremal states $\ket{\Psi_{k_i}}$ take the form
\begin{equation}
\label{eq:PCS_A+}
    \ket{\widetilde \Psi_\alpha^{\,k_i}}
    =\widetilde N(k_i,r)\sum_{n=0}^{N(k_i)}\dfrac{\alpha^n}{\widetilde G(k_i,n)}\ket{\Psi_{n+k_i}},
    \quad \text{for} \quad i=0,1,\ldots,l\quad,
\end{equation}
where $k_0\equiv0$, $k_{l+1}=+\infty$, $N(k_i)=k_{i+1}-k_i-1$, and the function $\widetilde G$ as well as the normalization constant $\widetilde N$ become  
\begin{equation}
    \widetilde G(k_i,n)=\dfrac{n!}{\sqrt{(n+k_i)!}\big[g(\mathcal{E}_{n+k_i})\big]!}, \qquad
    \widetilde N(k_i,r)=\Bigg(\sum_{n=0}^{N(k_i)}\dfrac{r^{2n}}{\widetilde G^2(k_i,n)}\Bigg)^{-\frac{1}{2}}.
\end{equation}

\subsection{General form of the bilayer graphene coherent states}

The bilayer graphene coherent states obtained in \cref{sec:girardello_CS,subsec:PCS} can be written in general as follows
\begin{equation}
\label{eq:coherent_states_general_form}
    \ket{\Psi_\alpha^\sk}=\mathcal{N}(\mk,r)\sum_{n=0}^{N}\dfrac{\alpha^n}{\mathcal{G}(\mk,n)}\ket{\Psi_{n+\sk}},
\end{equation}
where $\mk = k$ and $N = +\infty$ for the BGCS, while $\mk = k_i$ and $N = N(k_i) = k_{i+1} - k_i - 1$ for the GPCS. Additionally, $\mathcal{N} \in \{N, \widetilde{N}\}$ and $\mathcal{G} \in \{G, \widetilde{G}\}$, according to each case.

Let us note that if $g$ has no roots, and we take $g(\mathcal{E}_n) = 1$ $\forall\, n \in \mathbb{N}_0$, we get that $N(0, n) = \widetilde{N}(0, n) = e^{-r^2/2}$ and $G(0, n) = \widetilde{G}(0, n) = \sqrt{n!}$. This implies that both the GPCS $\ket{\widetilde{\Psi}_\alpha^{\,0}}$ and the BGCS $\ket{\Psi_\alpha^{\,0}}$ take the standard form
\begin{equation}
    \label{eq:coherent_states_standard_form}
    \ket{\Psi_\alpha^{\,0}}=e^{-r^2/2}\sum_{n=0}^{+\infty}\dfrac{\alpha^n}{\sqrt{n!}}\ket{\Psi_n}.
\end{equation}

\subsubsection{Transition probability, completeness relation, measure and inner product}

If $\mathbb{P}_n := \ket{\Psi_n}\bra{\Psi_n}$ is the projection operator for the eigenstate $\ket{\Psi_n}$ of bilayer graphene, the transition probability from the coherent state $\ket{\Psi_\alpha^{\sk}}$ to the eigenstate $\ket{\Psi_n}$ reads
\begin{equation}
\label{eq:transition_prob_k}
    P_n^\sk(\alpha)\equiv\left\langle\Psi_\alpha^\sk\left|
    \mathbb{P}_n\right
    |\Psi_\alpha^\sk\right\rangle 
    =\big|\braket{\Psi_n|\Psi_\alpha^\sk}\big|^2
    = \dfrac{\mathcal{N}^2(\mk,r)\,r^{2(n-\sk)}}{\mathcal{G}^2(\mk,n-\mk)},
    \quad n\in \mathcal{J}_{\sk},
\end{equation}
where $\mathcal{J}_k=\{k,k+1,\ldots\}$ for the BGCS and $\mathcal{J}_{k_i}=\{0,1,\ldots,N(k_i)\}$ for the GPCS. 
On the other hand, the completeness relation for the bilayer graphene coherent states takes the form
\begin{equation}
\label{eq:completeness_relation_general}
    \mathbb{1}_{\mathcal{H}_\sk}
    =\int \ket{\Psi_\alpha^\sk}\bra{\Psi_\alpha^\sk}
    \;\mathrm{d}\mu(\alpha)
    =\sum_{n=0}^{N}
    \frac{2\pi\ket{\Psi_{n+\sk}}\bra{\Psi_{n+\sk}}}{\mathcal{G}^2(\mk,n)}\int_0^{+\infty}\mathcal{N}^2(\mk,r)r^{2n+1}m(\mk,r)\,\mathrm{d}r,
\end{equation}
with $\mathbb{1}_{\mathcal{H}_\sk}$ being the identity operator in the subspace
\begin{equation}
    \mathcal{H}_\sk^{\sk+\sn}=\text{span}\Big(\{\ket{\Psi_{n+\sk}}\}_{n=0}^{N}\Big),
\end{equation}
and $m(\mk,r)$ being an unknown measure function. Thus, the problem of finding an appropriate measure $\mathrm{d}\mu(\alpha)$ to guarantee the validity of Eq.~\cref{eq:completeness_relation_general} is reduced to solve the integral equation
\begin{equation}
    \int_0^{+\infty}\mathcal{N}^2(\mk,r)r^{2n+1}m(\mk,r)\,\mathrm{d}r=\frac{\mathcal{G}^2(\mk,n)}{2\pi},
\end{equation}
for $m(\mk,r)$, which can be a complicated task that depends on the specific form of $\mathcal{N}(\mk,r)$ and $\mathcal{G}(\mk,r)$. However, once $m(\mk,r)$ is found the measure turns out to be
\begin{equation}
\label{eq:measure_general}
    \mathrm{d}\mu(\alpha)
    =m(\mk,r)\mathrm{d}^2\alpha,
\end{equation}
i.e., the bilayer graphene coherent states define an over-complete set in $\mathcal{H}_\sk^{\sk+\sn}$ for this measure. Meanwhile, the inner product between two coherent states with $\alpha=re^{i\theta}$ and $\alpha'=r'e^{i\theta'}$ becomes
\begin{equation}
\label{eq:internal_product_k}
    \braket{\Psi_\alpha^\sk|\Psi_{\alpha'}^\sk}
    =\dfrac{\mathcal{N}(\mk,r)\mathcal{N}(\mk,r')}{\mathcal{N}^2(\mk,\sqrt{rr'}e^{i(\theta'-\theta)/2})}.
\end{equation}
From this expression we can see that in the limit $r'\rightarrow r$ and $\theta'\rightarrow\theta$ the state $\ket{\Psi_{\alpha'}^\sk}$ converges to $\ket{\Psi_{\alpha}^\sk}$, since
\begin{equation}
    \lim_{\alpha'\rightarrow\alpha} \braket{\Psi_\alpha^\sk|\Psi_{\alpha'}^\sk}=1.
\end{equation}

In particular, for the coherent states of Eq.~\cref{eq:coherent_states_standard_form} we obtain the standard transition probability (Poisson distribution)
\begin{equation}
\label{eq:transition_prob_standard}
    P_n(\alpha)
    =\big|\braket{\Psi_n|\Psi_\alpha^{\,0}}\big|^2
    = \dfrac{r^{2n}}{n!}e^{-r^2},
    \quad n\in \mathbb{N}_0,
\end{equation}
and the standard measure with $m(0,r)=1/\pi$.

\subsubsection{Probability density and probability current}

According to Eq.~\cref{eq:density_BG}, the probability density for the bilayer graphene coherent states becomes 
\begin{equation}
    \rho_\alpha(r,\theta\,; x)
    :=\mathcal{N}^2(\mk,r)
    \sum_{n=0}^{N}\sum_{m=0}^{N} \dfrac{r^{n+m}}{\mathcal{G}(\mk,n)\mathcal{G}(\mk,m)} \cos{\big[(n-m)\theta\big]} \,\rho_{n,m}^\sk(x),
\end{equation}
where
\begin{equation}
\label{eq:rho_nm_general}
    \rho_{n,m}^\sk(x)
    = \Psi_{n+\sk}^\dagger(x) \Psi_{m+\sk}(x)
    = \dfrac{1}{2} \left(
    \psi_{n+\sk}^{(2)}(x) \psi_{m+\sk}^{(2)}(x)
    + \psi_{n+\sk}^{(0)}(x) \psi_{m+\sk}^{(0)}(x)
    \right).
\end{equation}
Moreover, from Eq.~\cref{eq:xcurrent_BG} the associated probability current in $x$-direction turns out to be
\begin{equation}
    J_{x,\alpha}(r,\theta\,; x)
    = \dfrac{\hbar}{2m^*}\, \mathcal{N}^2(\mk,r) 
    \sum_{n=0}^{N}\sum_{m=0}^{N}\dfrac{r^{n+m}}{\mathcal{G}(\mk,n)\mathcal{G}(\mk,m)} \sin{\big[(n-m)\theta\big]}\,M_{nm}^\sk(x),
\end{equation}
while in $y$-direction the probability current obtained from Eq.~\cref{eq:ycurrent_BG} becomes
\begin{equation}
    J_{y,\alpha}(r,\theta\,; x)
    = \dfrac{\hbar}{2m^*}\, \mathcal{N}^2(\mk,r) 
    \sum_{n=0}^{N}\sum_{m=0}^{N} \dfrac{r^{n+m}}{\mathcal{G}(\mk,n)\mathcal{G}(\mk,m)} \cos{\big[(n-m)\theta\big]}\,N_{nm}^\sk(x),
\end{equation}
where the functions $M_{nm}^\sk(x)$ and $N_{nm}^\sk(x)$ are defined by
\begin{equation}
    \begin{aligned}
    M_{nm}^\sk(x)
    &:=\psi_{n+\sk}^{(2)}(x) \left(\psi_{m+\sk}^{(0)}(x)\right)'
    + \psi_{n+\sk}^{(0)}(x) \left(\psi_{m+\sk}^{(2)}(x)\right)'\\[0.5em]
    &+ \frac{\eta(x)}{2}  \left(
    \psi_{n+\sk}^{(2)}(x) \psi_{m+\sk}^{(0)}(x)
    - \psi_{n+\sk}^{(0)}(x) \psi_{m+\sk}^{(2)}(x)
    \right),
\end{aligned}
\end{equation}
\begin{equation}
    \begin{aligned}
    N_{nm}^\sk(x)
    &:= \psi_{n+\sk}^{(0)}(x) \left(\psi_{m+\sk}^{(2)}(x)\right)'
    - \psi_{n+\sk}^{(2)}(x) \left(\psi_{m+\sk}^{(0)}(x)\right)'\\[0.5em]
    &- \dfrac{\eta(x)}{2}  \left(
    \psi_{n+\sk}^{(2)}(x) \psi_{m+\sk}^{(0)}(x)
    + \psi_{n+\sk}^{(0)}(x) \psi_{m+\sk}^{(2)}(x)
    \right).
\end{aligned}
\end{equation}
Note that for coherent states such that $\theta$ is an integer multiple of $\pi$ the current density in $x$-direction is zero, while for half-integer multiples of $\pi$ the current density in $y$-direction vanishes.

Having deduced the general form of the bilayer graphene coherent states and studied some of their properties, let us now consider the specific cases discussed in \cref{subsec:ladder_operators_bilayer_graphene}, which arise according to the roots of $f$.

\subsection{Case with \texorpdfstring{$f(\mathcal{E}_j)=f(\mathcal{E}_{j+1})=0$}{f(n)=0}}
\label{subsec:coherent_states_f_with_roots}

This condition corresponds to choose $\epsilon=\mathcal{E}_j$ in the diagonal case (both regular and critical) and in the non-diagonal regular case. Additionally, let us suppose that $f(\mathcal{E}_n) = 1\; \forall\, n \in \mathbb{N}_0 \backslash \mathcal{J}_+$.

\subsubsection{Barut-Girardello coherent states in \texorpdfstring{$\mathcal{H}_{j+1}^{\infty}$}{H j+1}}

Under these assumptions, the BGCS for bilayer graphene becomes
\begin{equation}
\label{eq:coherent_states_CI}
    \ket{\Psi_\alpha^{\,j+1}}
    =N(j+1,r)\sum_{n=0}^{+\infty}\frac{\alpha^n}{\sqrt{(n+j+1)!}}\ket{\Psi_{n+j+1}},
\end{equation}
where the normalization constant is given by
\begin{equation}
\label{eq:transition_prob_j+1}
    N(j+1,r)=e^{-r^2/2}r^{j+1}\left(1-e^{-r^2}\sum_{n=0}^{j}\frac{r^{2n}}{n!}\right)^{-\frac{1}{2}}.
\end{equation}
The probability transition from the coherent state $\ket{\Psi_\alpha^{\,j+1}}$ to the eigenstate $\ket{\Psi_n}$ is
\begin{equation}
\label{eq:transition_prob_H_k+1}
    P_n^{\,j+1}(\alpha)=N^2(j+1,r)\dfrac{r^{2(n-j-1)}}{n!} \quad \forall\;n\geq j+1.
\end{equation}
Moreover, the BGCS form an over-complete set in $\mathcal{H}_{j+1}^{\infty}$ with measure given by
\begin{equation}
\label{eq:measure_I}
    \mathrm{d}\mu(\alpha)
    =\frac{1}{\pi}\dfrac{r^{2j+2}}{N^2(j+1,r)}e^{-r^2}\,\mathrm{d}^2\alpha.
\end{equation}

Note that these expressions are valid in both cases \caseCI and \caseCII, i.e., for any $w_0 \in [0, +\infty)$. Nevertheless, although the extremal state is formally the same for both CS, the observable quantities (probability density, probability current, etc.) in each case differ significantly. This is due to the different properties of the SUSY transformation for each case, which affects the eigenfunctions of $H_2$ associated to the upper component of the bilayer graphene eigenstates. For instance, for $\epsilon = 2\omega$ in the regular case each component of the extremal state $\ket{\Psi_3}$ exhibits three nodes, since the SUSY transformation is isospectral for the two auxiliary Hamiltonians. However, in the critical case the lower component of $\ket{\Psi_3}$ has still three nodes but the upper component has only two, which is due to the well-known oscillation theorem and taking into account that in the critical case $\mathcal{E}_2 \notin \text{Sp}(H_2)$.

\begin{figure}[p]
    \centering

    \begin{subfigure}[b]{\textwidth}
        \centering
        {\small Probability density}
        \vspace{2em}
        
        \begin{subfigure}[b]{0.49\textwidth}
            \centering
            \includegraphics[width=\textwidth]{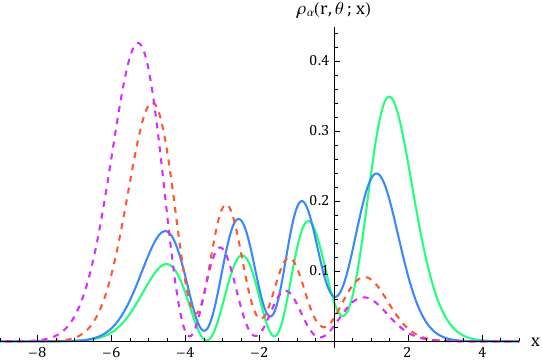}
            \caption{Regular case with $w_0=1$}
            \label{fig:density_BGCE_ϵ=2_w=1}
        \end{subfigure}
        \hfill
        \begin{subfigure}[b]{0.49\textwidth}
            \centering
            \includegraphics[width=\textwidth]{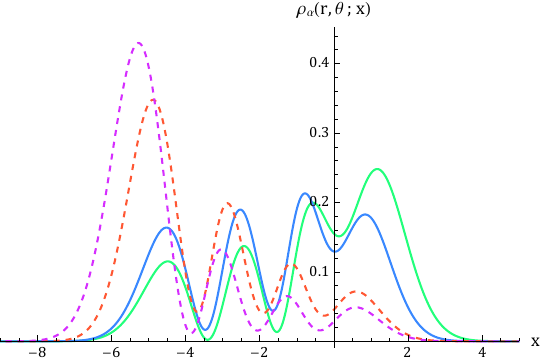}
            \caption{Critical case with $w_0=0$}
            \label{fig:density_BGCE_ϵ=2_w=0}
        \end{subfigure}
    \end{subfigure}
    
    \vspace{2em}

    \begin{subfigure}[b]{\textwidth}
        \centering
        {\small Probability current in $x$-direction}
        \vspace{2em}
        
        \begin{subfigure}[b]{0.49\textwidth}
            \centering
            \includegraphics[width=\textwidth]{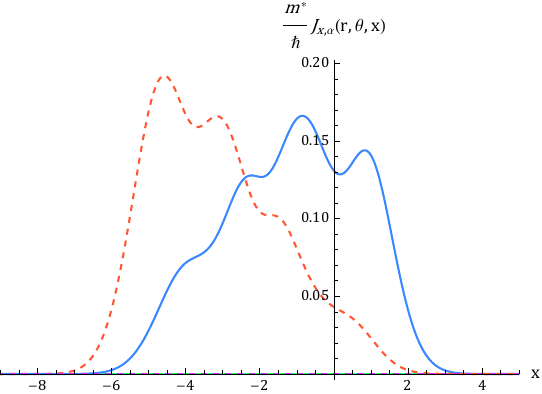}
            \caption{Regular case with $w_0=1$}
            \label{fig:currentX_BGCE_ϵ=2_w=1}
        \end{subfigure}
        \hfill
        \begin{subfigure}[b]{0.49\textwidth}
            \centering
            \includegraphics[width=\textwidth]{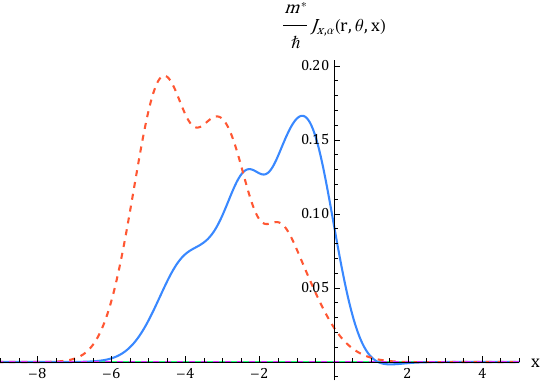}
            \caption{Critical case with $w_0=0$}
            \label{fig:currentX_BGCE_ϵ=2_w=0}
        \end{subfigure}
    \end{subfigure}
    
    \vspace{2em}

    \begin{subfigure}[b]{\textwidth}
        \centering
        {\small Probability current in $y$-direction}
        \vspace{2em}
        
        \begin{subfigure}[b]{0.49\textwidth}
            \centering
            \includegraphics[width=\textwidth]{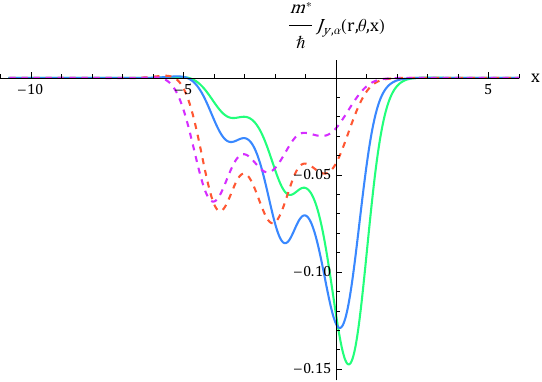}
            \caption{Regular case with $w_0=1$}
            \label{fig:currentY_BGCE_ϵ=2_w=1}
        \end{subfigure}
        \hfill
        \begin{subfigure}[b]{0.49\textwidth}
            \centering
            \includegraphics[width=\textwidth]{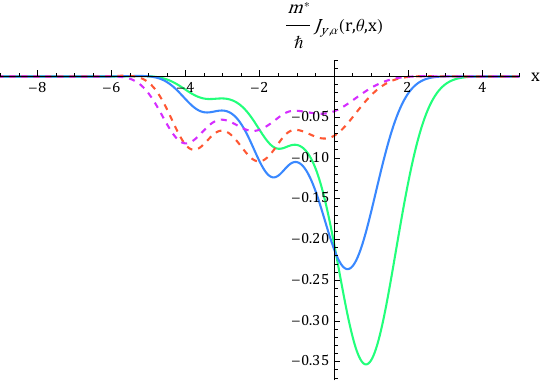}
            \caption{Critical case with $w_0=0$}
            \label{fig:currentY_BGCE_ϵ=2_w=0}
        \end{subfigure}
    \end{subfigure}

    \vspace{1em}
    \caption{BGCS for $\epsilon = 2\omega$ and fixed $r=1$, with $\theta=0$ (green), $\theta=\pi/3$ (blue), $\theta=2\pi/3$ (orange), and $\theta=\pi$ (purple).}
    \label{fig:partial_coherent_states}
\end{figure}

In \cref{fig:magnetic_field_2} the external magnetic field obtained from Eq.~\cref{eq:magnetic_field} for $\epsilon=2\omega$ is shown. The corresponding probability density and probability current for the BGCS of bilayer graphene are illustrated in \cref{fig:partial_coherent_states}.

\begin{figure}[p]
    \centering

    \begin{subfigure}[b]{\textwidth}
        \centering
        {\small Probability density}
        \vspace{2em}
        
        \begin{subfigure}[b]{0.49\textwidth}
            \centering
            \includegraphics[width=\textwidth]{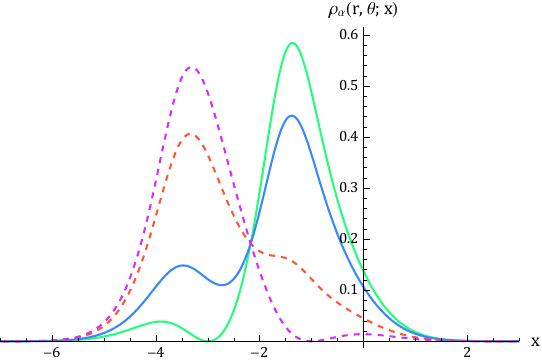}
            \caption{$r=1$}
            \label{fig:density_GPCS_ϵ=2_w=0_r=1}
        \end{subfigure}
        \hfill
        \begin{subfigure}[b]{0.49\textwidth}
            \centering
            \includegraphics[width=\textwidth]{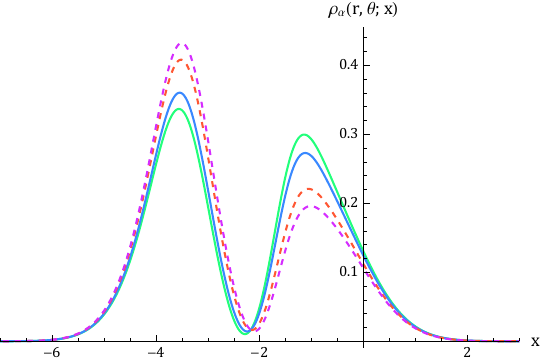}
            \caption{$r=10$}
            \label{fig:density_GPCS_ϵ=2_w=0_r=10}
        \end{subfigure}
    \end{subfigure}
    
    \vspace{2em}

    \begin{subfigure}[b]{\textwidth}
        \centering
        {\small Probability current in $x$-direction}
        \vspace{2em}
        
        \begin{subfigure}[b]{0.49\textwidth}
            \centering
            \includegraphics[width=\textwidth]{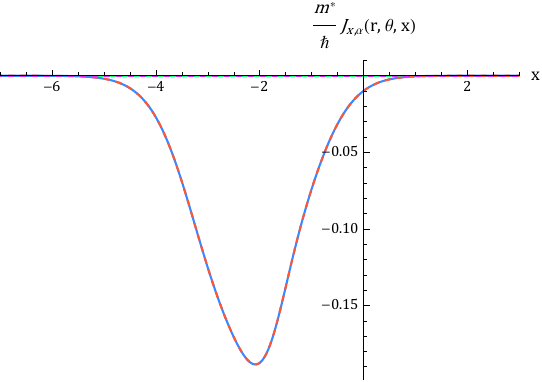}
            \caption{$r=1$}
            \label{fig:currentX_GPCS_ϵ=2_w=0_r=1}
        \end{subfigure}
        \hfill
        \begin{subfigure}[b]{0.49\textwidth}
            \centering
            \includegraphics[width=\textwidth]{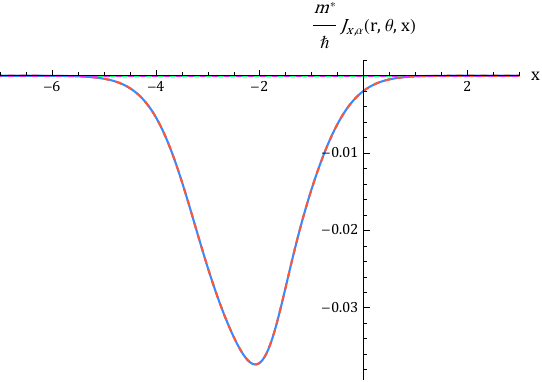}
            \caption{$r=10$}
            \label{fig:currentX_GPCS_ϵ=2_w=0_r=10}
        \end{subfigure}
    \end{subfigure}
    
    \vspace{2em}

    \begin{subfigure}[b]{\textwidth}
        \centering
        {\small Probability current in $y$-direction}
        \vspace{2em}
        
        \begin{subfigure}[b]{0.49\textwidth}
            \centering
            \includegraphics[width=\textwidth]{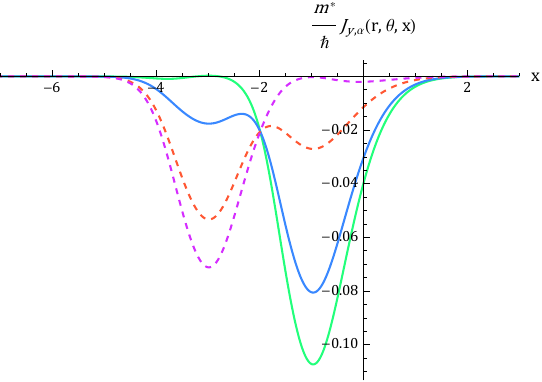}
            \caption{$r=1$}
            \label{fig:currentY_GPCE_ϵ=2_w=0_r=1}
        \end{subfigure}
        \hfill
        \begin{subfigure}[b]{0.49\textwidth}
            \centering
            \includegraphics[width=\textwidth]{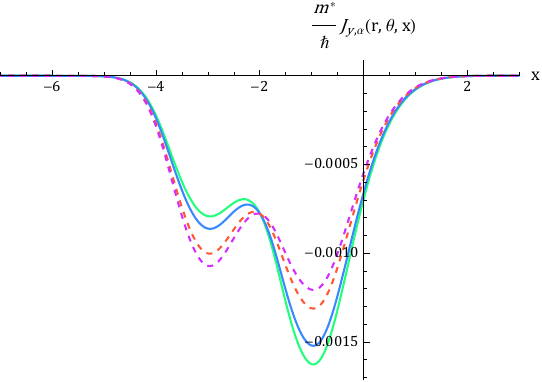}
            \caption{$r=10$}
            \label{fig:currentY_GPCE_ϵ=2_w=0_r=10}
        \end{subfigure}
    \end{subfigure}

    \vspace{1em}
    
    \caption{GPCS for $\epsilon = 2\omega$ and $w_0=0$ (critical case). The plots correspond to $r=1$ (left) and $r=10$ (right), with $\theta=0$ (green), $\theta=\pi/3$ (blue), $\theta=2\pi/3$ (orange), and $\theta=\pi$ (purple).}
    \label{fig:perelomov_coherent_states}
\end{figure}

\newpage
\subsubsection{Gilmore-Perelomov coherent states in \texorpdfstring{$\mathcal{H}_0^{j-1}$}{H0}}
\label{subsubsec:coherent_states_PCS}

The GPCS for bilayer graphene derived from the extremal state $\ket{\Psi_0}$ take the form
\begin{equation}
\label{eq:coherent_states_GP}
    \ket{\widetilde \Psi_\alpha^{\,j}}
    =\widetilde N(0,r)\sum_{n=0}^{j-1}\dfrac{\alpha^n}{\sqrt{n!}}\ket{\Psi_n},
\end{equation}
where the normalization constant and transition probability from $\ket{\widetilde \Psi_\alpha^{\,j}}$ to $\ket{\Psi_n}$ are 
\begin{equation}
\label{eq:transition_prob_H_0}
    \widetilde N(0,r)=\left(\sum_{n=0}^{j-1}\dfrac{r^{2n}}{n!}\right)^{-\frac{1}{2}}, \quad
    \widetilde P_n^{\,0}(\alpha)=\widetilde N^2(0,r)\dfrac{r^{2n}}{n!}\quad \forall\; n\leq j-1.
\end{equation}
Moreover, these coherent states define an overcomplete set in $\mathcal{H}_{0}^{j-1}$ with measure given by
\begin{equation}
    \mathrm{d}\mu(\alpha)=\frac{1}{\pi}\dfrac{e^{-r^2}}{\widetilde N^2(0,r)}\;\mathrm{d}^2\alpha.
\end{equation}

Contrasting with the BGCS of the preceding section, the observables associated with the GPCS exhibit a great similarity in both regular and critical cases. This happens since although the properties of each eigenstate of the set $\{\ket{\Psi_j}, \ket{\Psi_{j+1}}, \ldots\}$ are very different, the GPCS are just linear combinations of the first $j-1$ bilayer graphene eigenstates.

Figure \ref{fig:perelomov_coherent_states} shows the probability density and probability current for the GPCS for $\epsilon=2\omega$ and $w_0=0$ (critical case). Note that for $r=1$ such quantities differ significantly for each value of the phase $\theta$ taken. On the other hand, for $r=10$ they become very similar for the different values of $\theta$ chosen.

\subsection{Case with \texorpdfstring{$f(\mathcal{E}_n)\neq0$}{f(n) dif 0}}
\label{subsec:coherent_states_f_without_roots}

This condition can be imposed for $\epsilon\neq\mathcal{E}_j$, and in the non-diagonal critical case with $\epsilon=\mathcal{E}_j$. Let us consider as well a function $f$ such that $g(\mathcal{E}_n) = 1\, \forall\, n \in \mathbb{N}_0$, thus the BGCS and the GPCS coincide, taking the standard form given in Eq.~\cref{eq:coherent_states_standard_form} with the standard transition probability \cref{eq:transition_prob_standard}.

\subsubsection{Case with \texorpdfstring{$\epsilon \neq \mathcal{E}_j$}{epsilon different from Ej}}
\label{subsubsec:coherent_states_CIV}

\begin{figure}[p]
    \centering
    
    \begin{subfigure}[b]{\textwidth}
        \centering
        {\small Probability density}
        \vspace{2em}
        
        \begin{subfigure}[b]{0.49\textwidth}
            \centering
            \includegraphics[width=\textwidth]{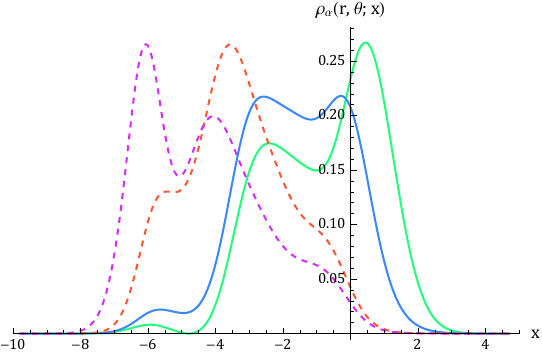}
            \caption{Regular case with $w_0=1$}
            \label{fig:density_BGCE_w=1}
        \end{subfigure}
        \hfill
        \begin{subfigure}[b]{0.49\textwidth}
            \centering
            \includegraphics[width=\textwidth]{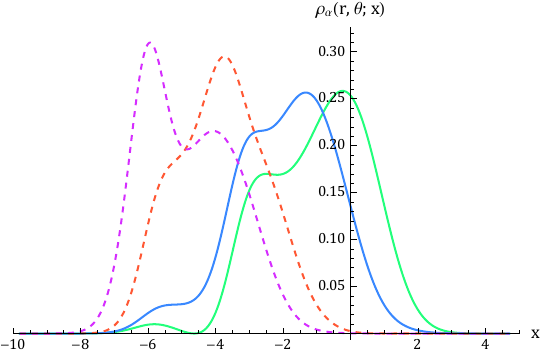}
            \caption{Critical case with $w_0=0$}
            \label{fig:density_BGCE_w=0}
        \end{subfigure}
    \end{subfigure}
    
    \vspace{2em}

    \begin{subfigure}[b]{\textwidth}
        \centering
        {\small Probability current in $x$-direction}
        \vspace{2em}
        
        \begin{subfigure}[b]{0.49\textwidth}
            \centering
            \includegraphics[width=\textwidth]{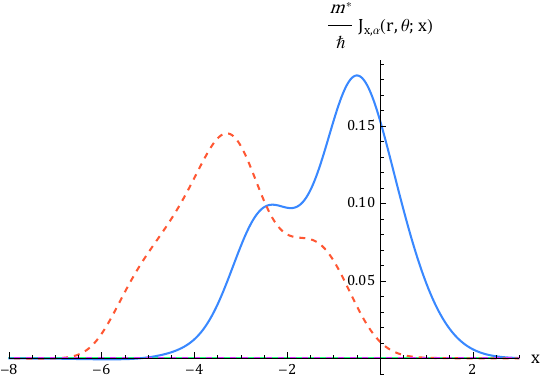}
            \caption{Regular case with $w_0=1$}
            \label{fig:currentX_BGCE_w=1}
        \end{subfigure}
        \hfill
        \begin{subfigure}[b]{0.49\textwidth}
            \centering
            \includegraphics[width=\textwidth]{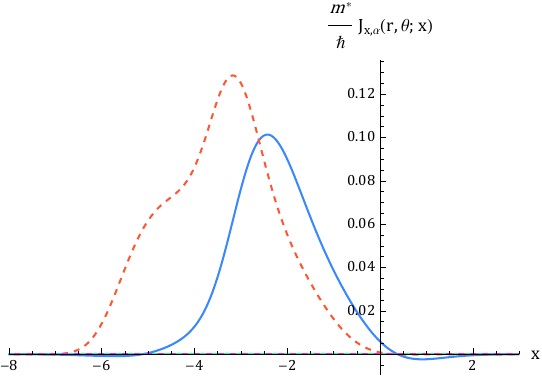}
            \caption{Critical case with $w_0=0$}
            \label{fig:currentX_BGCE_w=0}
        \end{subfigure}
    \end{subfigure}
    
    \vspace{2em}

    \begin{subfigure}[b]{\textwidth}
        \centering
        {\small Probability current in $y$-direction}
        \vspace{2em}
        
        \begin{subfigure}[b]{0.49\textwidth}
            \centering
            \includegraphics[width=\textwidth]{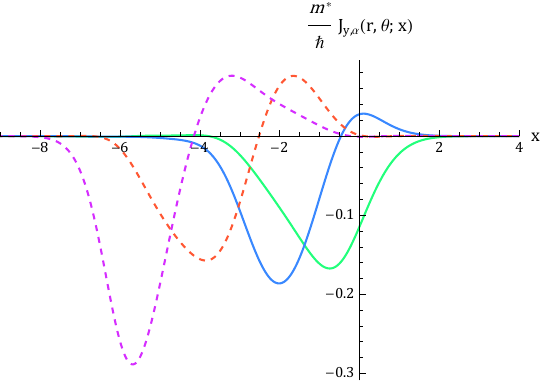}
            \caption{Regular case with $w_0=1$}
            \label{fig:currentY_BGCE_w=1}
        \end{subfigure}
        \hfill
        \begin{subfigure}[b]{0.49\textwidth}
            \centering
            \includegraphics[width=\textwidth]{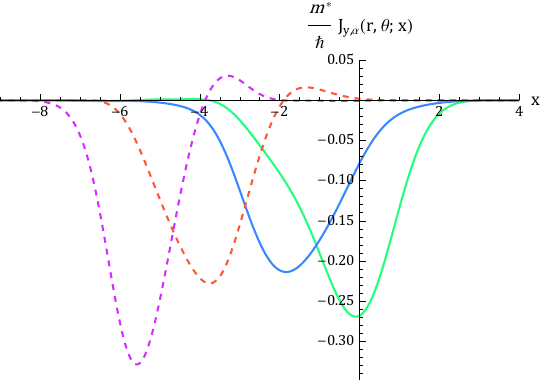}
            \caption{Critical case with $w_0=0$}
            \label{fig:currentY_BGCE_w=0}
        \end{subfigure}
    \end{subfigure}

    \vspace{2em}
    
    \caption{Bilayer graphene coherent states for $\epsilon = \omega/2$, $r=1$ and $\theta=0$ (green), $\theta=\pi/3$ (blue), $\theta=2\pi/3$ (orange), and $\theta=\pi$ (purple).}
    \label{fig:coherent_states_standard}
\end{figure}

In this case the completeness relation for the bilayer graphene coherent states reads 
\begin{equation}
\label{eq:completeness_relation_standard}
    \frac{1}{\pi} \int \ket{\Psi_\alpha}\bra{\Psi_\alpha}
    \;\mathrm{d}^2\alpha    =\sum_{n=0}^{+\infty}\ket{\Psi_n}\bra{\Psi_n}
    =\mathds{1}-\mathbb{P}_\nu,
\end{equation}
where $\mathds{1}$ is the identity operator in the full state space of bilayer graphene. Note that in the critical case $\mathbb{P}_\nu$ becomes the null operator, i.e., these states define an over-complete set in the full state space in such a case. In \cref{fig:coherent_states_standard} the probability density and probability currents for the bilayer graphene CS with $\epsilon=1/2$ are shown.

\subsubsection{Case with \texorpdfstring{$\epsilon = \mathcal{E}_j$}{epsilon = Ej}}
\label{subsubsec:coherent_states_critical_case}

Choosing now $f(\mathcal{E}_j) = f(\mathcal{E}_{j+1}) = \sqrt{2}$ and $f(\mathcal{E}_n) = 1\, \forall\, n \in \mathbb{N}_0 \backslash \mathcal{J}_+$ we can ensure that $g(\mathcal{E}_n) = 1$ $\forall\,n \in \mathbb{N}_0$. In this case the completeness relation takes the same form as for the standard coherent states, i.e., the bilayer graphene coherent states form an overcomplete set in the full state space.

\begin{figure}[p]
    \centering
    
    \begin{subfigure}[b]{\textwidth}
        \centering
        {\small Probability density}
        \vspace{2em}
        
        \begin{subfigure}[b]{0.49\textwidth}
            \centering
            \includegraphics[width=\textwidth]{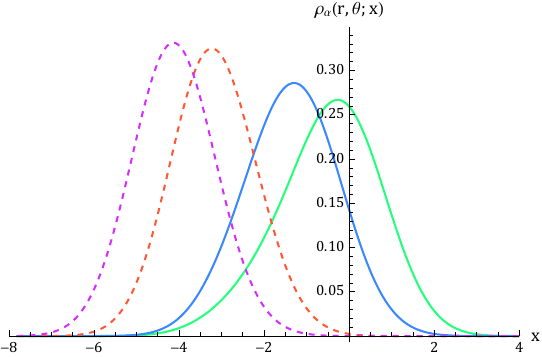}
            \caption{$r=1$}
            \label{fig:density_BGCE_r=1_nd}
        \end{subfigure}
        \hfill
        \begin{subfigure}[b]{0.49\textwidth}
            \centering
            \includegraphics[width=\textwidth]{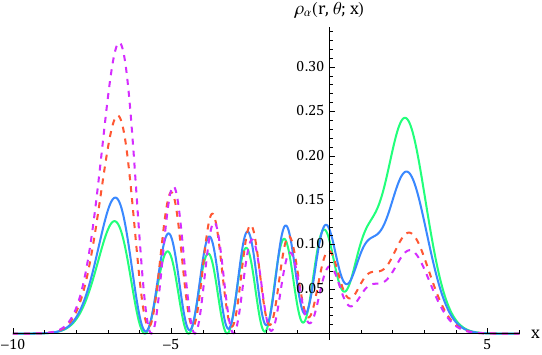}
            \caption{$r=10$}
            \label{fig:density_BGCE_r=10_nd}
        \end{subfigure}
    \end{subfigure}
    
    \vspace{2em}

    \begin{subfigure}[b]{\textwidth}
        \centering
        {\small Probability current in $x$-direction}
        \vspace{2em}
        
        \begin{subfigure}[b]{0.49\textwidth}
            \centering
            \includegraphics[width=\textwidth]{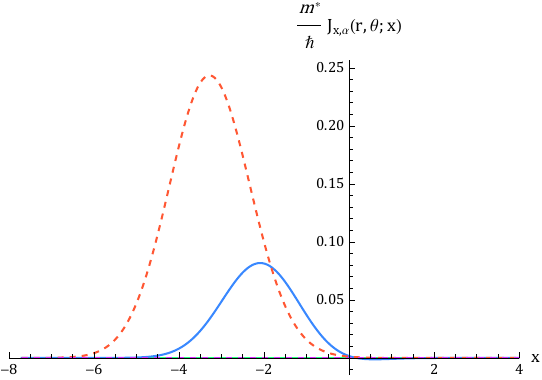}
            \caption{$r=1$}
            \label{fig:currentX_BGCE_r=1_nd}
        \end{subfigure}
        \hfill
        \begin{subfigure}[b]{0.49\textwidth}
            \centering
            \includegraphics[width=\textwidth]{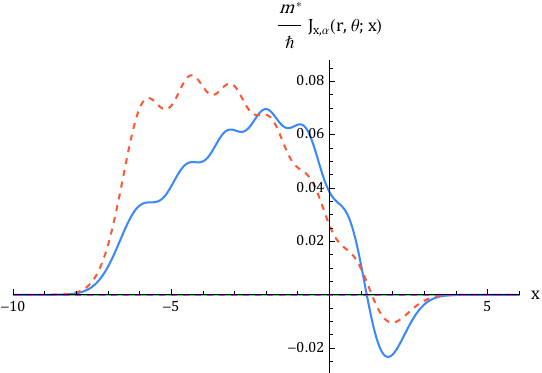}
            \caption{$r=10$}
            \label{fig:currentX_BGCE_r=10_nd}
        \end{subfigure}
    \end{subfigure}
    
    \vspace{2em}

    \begin{subfigure}[b]{\textwidth}
        \centering
        {\small Probability current in $y$-direction}
        \vspace{2em}
        
        \begin{subfigure}[b]{0.49\textwidth}
            \centering
            \includegraphics[width=\textwidth]{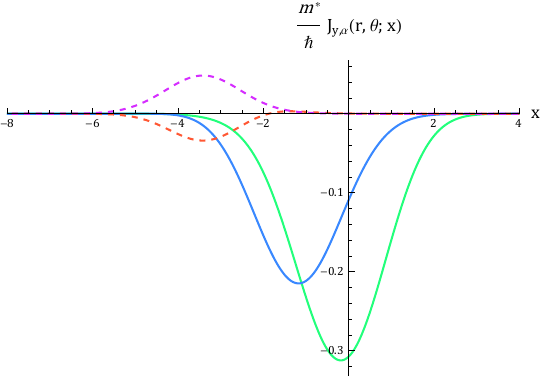}
            \caption{$r=1$}
            \label{fig:currentY_BGCE_r=1_nd}
        \end{subfigure}
        \hfill
        \begin{subfigure}[b]{0.49\textwidth}
            \centering
            \includegraphics[width=\textwidth]{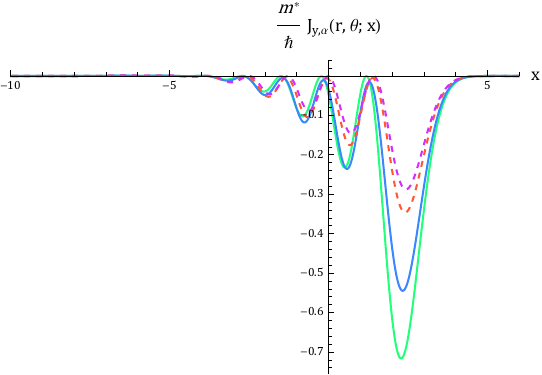}
            \caption{$r=10$}
            \label{fig:currentY_BGCE_r=10_nd}
        \end{subfigure}
    \end{subfigure}

    \vspace{2em}
    
    \caption{Bilayer graphene coherent states for $\epsilon=0$ and $w_0=0$ in the non-diagonal case. We have taken $r=1$ (left) and $r=10$ (right), with $\theta=0$ (green), $\theta=\pi/3$ (blue), $\theta=2\pi/3$ (orange), and $\theta=\pi$ (purple).}
    \label{fig:coherent_states_non_diag}
\end{figure}

Figure \ref{fig:coherent_states_non_diag} shows the probability density and probability current for the bilayer graphene coherent states in the critical case ($w_0=0$). For $r=1$ we can see that the probability density resembles a Gaussian, indicating that the main contribution comes from the extremal state $\ket{\Psi_0}$. However, for $r=10$ the contributions from other eigenstates start to appear, as the density shows both maxima and minima. Additionally, as $r$ increases the probability density and probability current tend to certain functions depending just on $x$, independently of the values of $\theta$ taken, except for those $\theta$ values where $\rho$ and $J$ vanish.

Finally, let us stress that although in the cases considered in this section it is not required to choose the function $f$ having roots at $\mathcal{E}_j$ and $\mathcal{E}_{j+1}$, this choice still can be done. Furthermore, if we choose $g(\mathcal{E}_n) = 1 \; \forall\, n \in \mathbb{N}_0 \backslash \mathcal{J}_+$ the BGCS take the form \cref{eq:coherent_states_CI}, while the GPCS in $\mathcal{H}_0^{j-1}$ are given by Eq.~\cref{eq:coherent_states_GP}.


\section{Time evolution for the bilayer graphene coherent states}
\label{sec:time_evolution}

The time evolution of the bilayer graphene coherent states \cref{eq:coherent_states_general_form} is obtained by applying the time evolution operator $\displaystyle U(t, t_0) = e^{-i \mathbb{H} (t - t_0) / \hbar}$ to the initial coherent state $\ket{\Psi_\alpha^\sk(t_0)}$. Taking into account the spectrum \cref{eq:spectrum_bilayer_graphene} of bilayer graphene, and fixing the initial time at $t_0 = 0$, we have that
\begin{equation}
 \label{eq:time_evolution_general}
    \begin{aligned}
        \ket{\Psi_\alpha^\sk(t)}
        &=U(t,0) \ket{\Psi_\alpha^\sk(0)}
        =e^{-i\mathbb{H}t/\hbar}\ket{\Psi_\alpha^\sk(0)}
        =\mathcal{N}(\mk,r)\sum_{n=0}^{N}\dfrac{\alpha^n e^{-i\omega_0\abs{n+\sk-\nu}t}}{\mathcal{G}(\mk,n)}\ket{\Psi_{n+\sk}}\\[1em]
        &=\mathcal{N}(\mk,r)\left(e^{-i(\nu-\sk)\omega_0 t}\sum_{n=0}^{k-1}\frac{\alpha^n e^{in\omega_0 t}}{\mathcal{G}(\mk,n)}\ket{\Psi_{n+\sk}}
        +e^{i(\nu-\sk)\omega_0 t}\sum_{n=k}^{N}\dfrac{\alpha^n e^{-in\omega_0 t}}{\mathcal{G}(\mk,n)}\ket{\Psi_{n+\sk}}\right),
    \end{aligned}
\end{equation}
where
\begin{equation}
    k = \left\{
    \begin{array}{cc}
         \lceil \nu-\mk \rceil & \quad \nu \geq \mk \\[0.5em]
         0 & \quad \nu < \mk
    \end{array},
    \right.
\end{equation}
with $\lceil x \rceil$ representing the ceiling function of $x$. Thus, the time-dependent probability density becomes
\begin{equation}
\label{eq:time_density_general}
\begin{aligned}
    \rho_\alpha(r,\theta\,; x, t)
    =\mathcal{N}^2(\mk,r)\sum_{n=0}^{N}\sum_{m=0}^{N}\dfrac{r^{n+m}}{\mathcal{G}(\mk,n)\mathcal{G}(\mk,m)}
    \cos\Big(\big(m-n\big)\theta- \big(\abs{m+\mk-\nu}-\abs{n+\mk-\nu}\big)\omega_0 t\Big)\rho_{n,m}^\sk(x),
\end{aligned}
\end{equation}
where $\rho_{n,m}^\sk$ is given by Eq.~\cref{eq:rho_nm_general}. In particular, for $\nu < \mk$ the time evolution in \cref{eq:time_evolution_general,eq:time_density_general} reduces to
\begin{equation}
 \label{eq:coherent_evolution_GCS}
    \ket{\Psi_\alpha^\sk(t)}
        =e^{i(\nu-\sk)\omega_0 t}\ket{\Psi_{\alpha(t)}^\sk}, \qquad \rho_\alpha(r,\theta\,; x, t)=\rho_\alpha(r,\theta\,-\omega_0 t; x),
\end{equation}
with $\alpha(t) = \alpha e^{-i\omega_0 t}$. Likewise, for $N<k$ it follows that
\begin{equation}
\label{eq:coherent_evolution_PCS}
    \ket{\Psi_\alpha^\sk(t)}
    =e^{-i(\nu-\sk)\omega_0 t}\ket{\Psi_{\alpha(t)}^\sk}, \qquad \rho_\alpha(r,\theta\,; x, t)=\rho_\alpha(r,\theta+\omega_0 t; x),
\end{equation}
where now $\alpha(t)=\alpha e^{i\omega_0 t}$. In both cases the bilayer graphene coherent states exhibit time coherence or temporal stability, i.e., the initial coherent state evolves into another coherent state at any time. Moreover, they evolve cyclically with period $\tau=2\pi/\omega_0$ and global phase $\varphi=\pm2\pi(\nu-\mk)$.

Let us analyze as well the fidelity $\displaystyle F(\Psi_\alpha(t_0),\Psi_\alpha(t)):=\abs{\braket{\Psi_\alpha(t_0)|\Psi_\alpha(t)}}^2$ of the coherent states, which measures the similarity between the state at time $t$ and the initial state at $t_0$. For the coherent states \cref{eq:time_evolution_general} such a fidelity becomes (we take $t_0=0$):
\begin{equation}
\begin{aligned}
\label{eq:fidelity_arbitrary}
    F\left(\Psi_\alpha(0),\Psi_\alpha(t)\right)
    =\mathcal{N}^4(\mk,r)&\Bigg[\left(\sum_{n=0}^{+\infty}\dfrac{r^{2n}}{\mathcal{G}(\mk,n)}\cos{((n+\mk-\nu)\omega_0t)}\right)^2\\
    &+\left(\sum_{n=0}^{+\infty}\dfrac{r^{2n}}{\mathcal{G}(\mk,n)}\sin{((n+\mk-\nu)\omega_0t)}\right)^2\Bigg],
\end{aligned}
\end{equation}
which is independent of $\theta$ and $w_0$, since it does not depend on the explicit form of the eigenstates $\ket{\Psi_n}$.

Let us analyze next the time evolution of the coherent states obtained in  \cref{subsec:coherent_states_f_with_roots,subsec:coherent_states_f_without_roots}.

\begin{figure}[bt]
    \centering
    \begin{subfigure}[b]{0.49\textwidth}
        \centering
        \includegraphics[width=\textwidth]{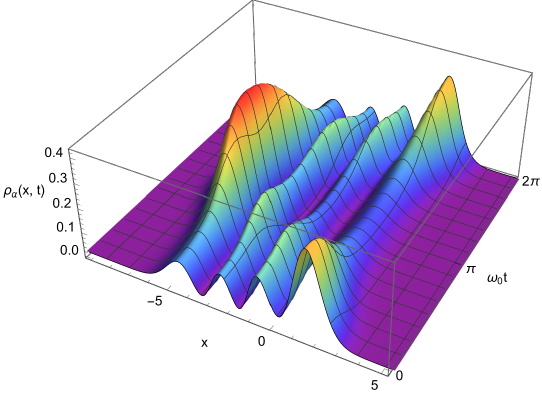}
        \caption{Regular case with $w_0=1$}
        \label{fig:densityt_ϵ=2_w0=1}
    \end{subfigure}
    \hfill
    \begin{subfigure}[b]{0.49\textwidth}
        \centering
        \includegraphics[width=\textwidth]{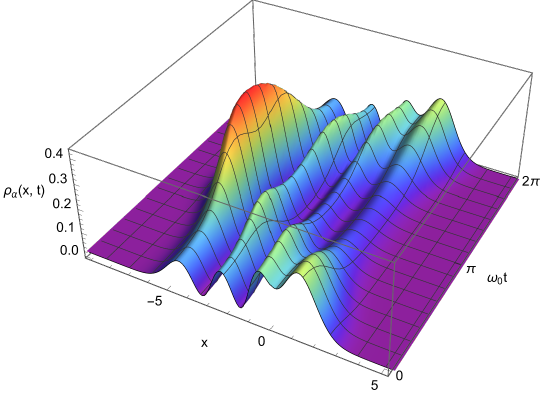}
        \caption{Critical case with $w_0=0$}
        \label{fig:densityt_ϵ=2_w0=0}
    \end{subfigure}
    \caption{Time-dependent probability density for the Barut-Girardello CS with $\epsilon=2\omega$, $r=1$ and $\theta=0$. These states exhibit coherent and cyclic time evolution with period $\tau=2\pi/\omega_0$.}
    \label{fig:densities_temporal_evolution_CIII}
\end{figure}

\subsection{Case with \texorpdfstring{$f(\mathcal{E}_j)=f(\mathcal{E}_{j+1})=0$}{f(n)=0}}
\label{subsec:time_evolution_f_with_roots}

\subsubsection{Barut-Girardello coherent states in \texorpdfstring{$\mathcal{H}_{j+1}^{\infty}$}{H j+1}}
\label{subsubsec:time_evolution_GCS}

We have in this case that $\nu = j < \mk = j + 1$, thus the BGCS evolve according to Eq.~\cref{eq:coherent_evolution_GCS}, exhibit temporal stability and evolve cyclically with period $\tau = 2\pi / \omega_0$ and global phase $\varphi = -2\pi$. Additionally, the geometric phase (see Eq.~\cref{eq:beta_t_indep} of the \ref{sec:cyclic_evolution}) takes the form 
\begin{equation}
\label{eq:geometric_phase_I}
    \beta
    = 2\pi\Big( r^2 + (j+1)P_{j+1}^{\,j+1}(\alpha) \Big),
\end{equation}
where $P_{j+1}^{\,j+1}(\alpha)$ is the transition probability of Eq.~\cref{eq:transition_prob_H_k+1} with $n=j+1$.
Moreover, as the second term of Eq.~\cref{eq:geometric_phase_I} vanishes for large amplitudes, the geometric phase tends to the one of the standard coherent states in such a limit.

\begin{figure}[bt]
    \centering
    \begin{subfigure}[b]{0.49\textwidth}
        \centering
        \includegraphics[width=\textwidth]{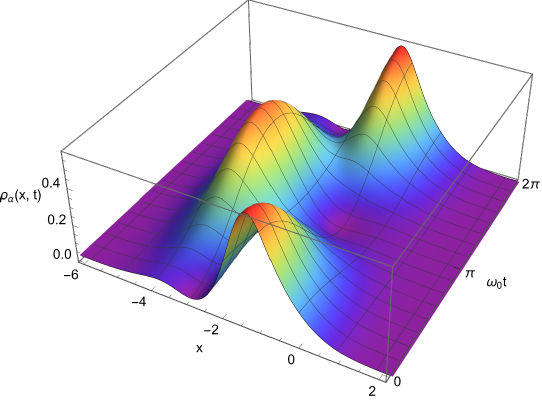}
        \caption{Case with $r=1$}
        \label{fig:densityt_ϵ=2_w0=0_r=1}
    \end{subfigure}
    \hfill
    \begin{subfigure}[b]{0.49\textwidth}
        \centering
        \includegraphics[width=\textwidth]{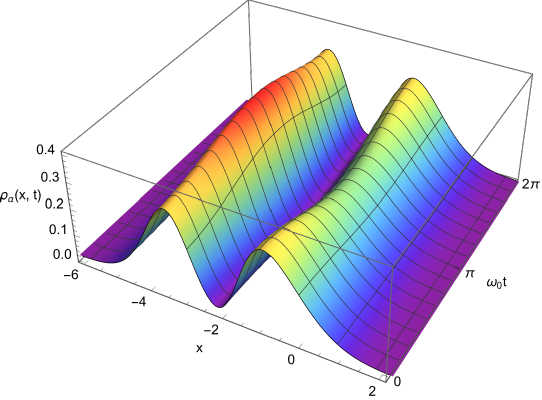}
        \caption{Case with $r=10$}
        \label{fig:densityt_ϵ=2_w0=0_r=10}
    \end{subfigure}
    \caption{Time-dependent probability density for the Gilmore-Perelomov CS with $\epsilon=2\omega$ and $\theta=0$. These states show coherent and cyclic time evolution with period $\tau=2\pi/\omega_0$.}
    \label{fig:densities_temporal_evolution_GP_CI}
\end{figure}

Figure \ref{fig:densities_temporal_evolution_CIII} shows the time-dependent probability density for the Barut-Girardello CS with $\epsilon=2\omega$.

\subsubsection{Gilmore-Perelomov coherent states in \texorpdfstring{$\mathcal{H}_0^{j-1}$}{H0}}
\label{subsubsec:time_evolution_PCS}

In this case we have that $\nu = j$ and $\mk = 0$, so that $N = j - 1 < k = j$. Thus, the GPCS evolve as in Eq.~\cref{eq:coherent_evolution_PCS}, exhibit temporal coherence and evolve cyclically with period $\tau = 2\pi / \omega_0$ and global phase $\varphi = -2\pi j$. Moreover, for these states the geometric phase becomes
\begin{equation}
    \beta
    =-2\pi\sum_{n=0}^{j-1}n\widetilde P_n^{\,j}(\alpha)
    \equiv-2\pi\langle n \rangle_\alpha,
\end{equation}
where $\widetilde P_n^{\,j}(\alpha)$ is the transition probability \cref{eq:transition_prob_H_0}.

Figure \ref{fig:densities_temporal_evolution_GP_CI} illustrates the associated time-dependent probability density for $\epsilon=2\omega$. We observe that for $r=10$, the probability density remains almost constant over time. 

\subsection{Case with \texorpdfstring{$f(\mathcal{E}_n)\neq0$}{f(n) dif 0}}
\label{subsec:time_evolution_f_without_roots}

\subsubsection{Case with \texorpdfstring{$\epsilon \neq \mathcal{E}_j$}{epsilon different from Ej}}
\label{subsubsec:time_evolution_CIV}

Now $\mk=0$, thus for $\nu < 0$ the bilayer graphene coherent states evolve coherently and cyclically, with a period $\tau = 2\pi/\omega_0$, a global phase $\varphi = 2\pi\nu$, and a geometric phase $\beta = 2\pi r^2$, which coincides with the standard coherent states result.

On the other hand, for $\nu > 0$ the time evolution in general is not coherent. However, when $\epsilon$ is a rational multiple of $\omega$ such that $2\epsilon=(p/q)\omega$, with $p, q \in \mathbb{N}$ being coprime, the coherent states evolves cyclically with a period $\tau=2\pi q/\omega_0$ and $\varphi=\pi p$. The geometric phase for these states becomes
\begin{equation}
\label{eq:geometric_phase_arbitrary}
    \beta = 2\pi q \left[(r^2-\nu)
         \left(1-2\sum_{n=0}^{k-1}P_n(\alpha)
         \right)+2kP_k(\alpha)\right].
\end{equation}

Let us note that since the rational numbers are \textit{dense} in the real numbers, if $\epsilon$ is an irrational multiple of $\omega$ we can always find another factorization energy that is a rational multiple of $\omega$ as close as desired to the first one. This implies that we can always find a time period $\tau$ such that the bilayer graphene coherent state at that time becomes as close as we want to the initial state, up to a global phase.

\begin{figure}[bt]
    \centering
    \begin{subfigure}[b]{0.49\textwidth}
        \centering
        \includegraphics[width=\textwidth]{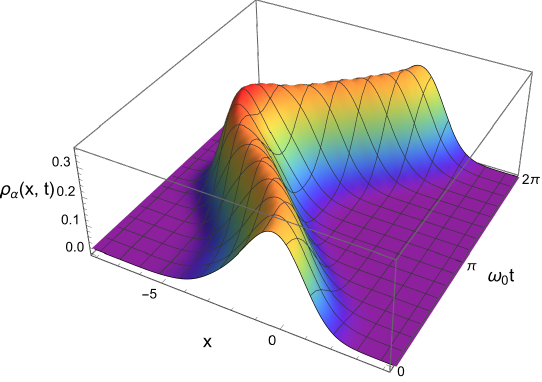}
        \caption{Critical case with $\epsilon=-\omega$: coherent cyclic time evolution with period $\tau=2\pi/\omega_0$.}
        \label{fig:densityt_ϵ=-1_w0=0}
    \end{subfigure}
    \hfill
    \begin{subfigure}[b]{0.49\textwidth}
        \centering
        \includegraphics[width=\textwidth]{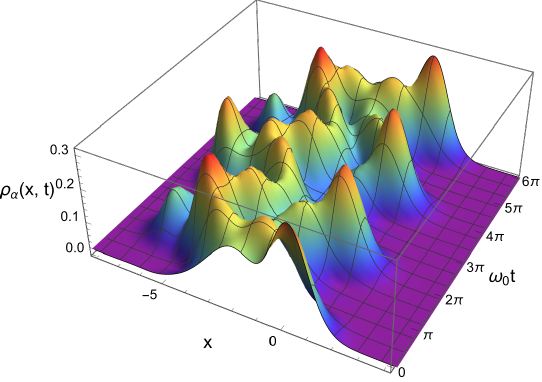}
        \caption{Regular case with $w_0=1$ and $\epsilon=\omega/3$: non-coherent cyclic time evolution with period $\tau=6\pi/\omega_0$.}
        \label{fig:densityt_ϵ=0.333_w0=1}
    \end{subfigure}
    \caption{Time-dependent probability density for the bilayer graphene coherent states with $r=1$, $\theta=0$, and two different factorization energies.}
    \label{fig:densities_temporal_evolution}
\end{figure}

\begin{figure}[bt]
    \centering
    \centering
    \includegraphics[scale=1.7]{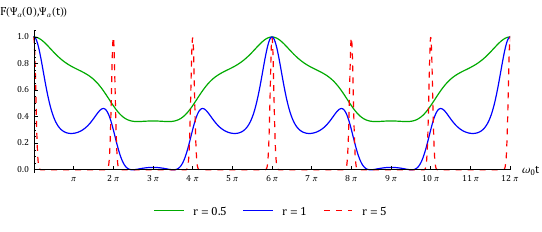}
    \caption{Fidelity as function of time for the bilayer graphene coherent states with $\epsilon=\omega/3$.}
    \label{fig:fidelity_ϵ=1/3}
\end{figure}

The time-dependent probability density for the bilayer graphene coherent states with $r=1$ and $\theta=0$ is shown in \cref{fig:densities_temporal_evolution}. In \cref{fig:densityt_ϵ=-1_w0=0} such evolution is coherent and cyclic, with period $\tau=2\pi/\omega_0$ for $\epsilon=-\omega$ and $w_0=0$. On the other hand, the evolution in \cref{fig:densityt_ϵ=0.333_w0=1} is not coherent, although it is still cyclic with period $\tau=6\pi/\omega_0$ for $\epsilon=\omega/3$ and $w_0=1$. For this latter case the fidelity as function of time can be seen in \cref{fig:fidelity_ϵ=1/3} for different values of $r$. We observe that the coherent states with $r=5$, in addition of having the period $\tau=6\pi/\omega_0$ tend to return to the same state (up to a global phase) for a different period $\tilde{\tau}=2\pi/\omega_0$.

\subsubsection{Case with \texorpdfstring{$\epsilon = \mathcal{E}_j$}{epsilon = Ej}}
\label{subsubsec:time_evolution_critical_case}

The time evolution for these coherent states can be obtained as a particular case of the previous results with $p=2j$ and $q=1$. This means that the time evolution is not coherent, but it is cyclic with period $\tau = 2\pi / \omega_0$ and global phase $\varphi = 2\pi j$. The associated geometric phase arises from Eq.~\cref{eq:geometric_phase_arbitrary} for $\nu = j$.
Note that for $j = 0$ the geometric phase coincides with the one for the standard coherent states. Additionally, the fidelity is obtained from Eq.~\cref{eq:fidelity_arbitrary} by setting $\nu = j$, from which it is clear that for $\tau = 2\pi / \omega_0$ such a fidelity is equal to $1$, regardless the value of $r$.

Finally, let us analyze the uncertainty product of a pair of quadratures for the bilayer graphene coherent states.

\section{Commutation relation and uncertainty product}
\label{sec:commutation_relation}

The two natural quadratures associated with the bilayer graphene ladder operators are defined by
\begin{equation}
    \mathbb{X}:=\dfrac{1}{\sqrt{2}}\left(\mathbb{A}^-+\mathbb{A}^+\right),
    \quad
    \mathbb{P}:=\dfrac{1}{i\sqrt{2}}\left(\mathbb{A}^--\mathbb{A}^+\right),
\end{equation}
which implies that
\begin{equation}
\label{eq:commutation_relation_X_A}
    \left[\mathbb{X},\mathbb{P}\right]
    =i\left[\mathbb{A}^-,\mathbb{A}^+\right].
\end{equation}
Moreover, it is known that the Barut-Girardello coherent states saturate the uncertainty product of such quadratures \cite{Fernandez2022_GGCoherentStates}, which in general fulfill
\begin{equation}
\label{eq:uncertainties_product_X_P}
    \Delta\mathbb{X}\Delta\mathbb{P}
    \geq
    \dfrac{1}{2}\left|\left\langle\left[\mathbb{X},
    \mathbb{P}\right]\right\rangle\right|.
\end{equation}
From Eq.~\cref{eq:commutation_relation_X_A} we deduce that the commutator of the quadratures is given by the action of the ladder operators onto the energy eigenstates, which in turn is determined by the roots of $f$.

Next, we will analyze the uncertainty product \cref{eq:uncertainties_product_X_P} for the coherent states of \cref{subsec:coherent_states_f_with_roots,subsec:coherent_states_f_without_roots}. Recall that if $\mathcal{E}_n$ is not a root of $g$, then $g(\mathcal{E}_n) = 1$ in these cases.

\begin{figure}[bt]
    \centering
    \begin{subfigure}[b]{0.49\textwidth}
        \centering
        \includegraphics[width=\textwidth]{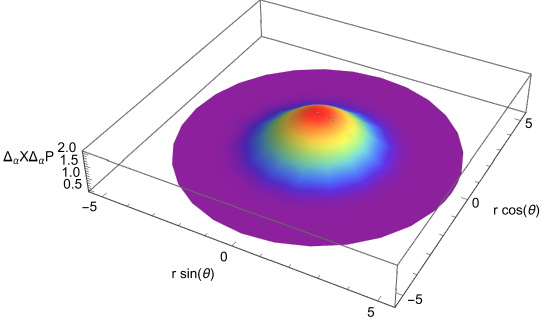}
        \caption{$j=2$}
        \label{fig:uncert_product_k=2}
    \end{subfigure}
    \hfill
    \begin{subfigure}[b]{0.49\textwidth}
        \centering
        \includegraphics[width=\textwidth]{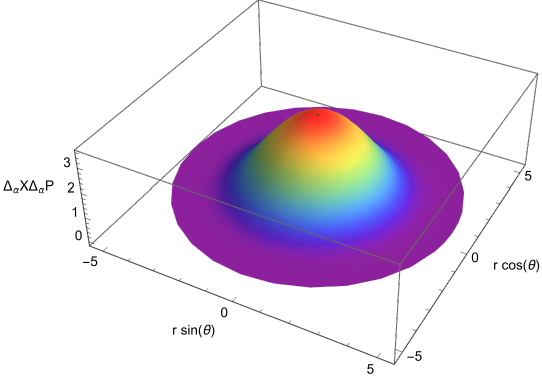}
        \caption{$j=5$}
        \label{fig:uncert_product_k=3}
    \end{subfigure}
    \caption{Uncertainty product of the quadratures for the BGCS with $g(\mathcal{E}_{j+1})=0$ and different values of $j$.}
    \label{fig:uncert_product_BG}
\end{figure}

\begin{figure}[bt]
    \centering
    \begin{subfigure}[b]{0.49\textwidth}
        \centering
        \includegraphics[width=\textwidth]{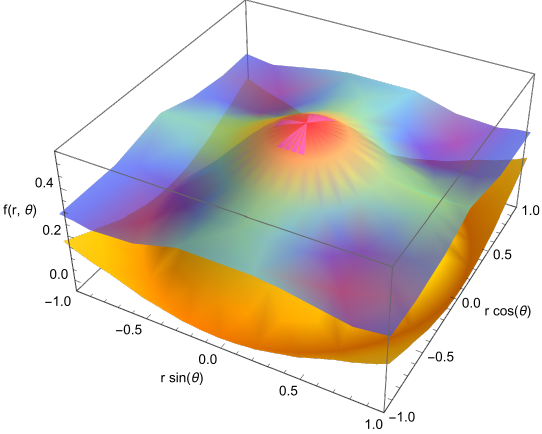}
        \caption{Close-up view}
        \label{fig:uncert_product_j=2_v1}
    \end{subfigure}
    \hfill
    \begin{subfigure}[b]{0.49\textwidth}
        \centering
        \includegraphics[width=\textwidth]{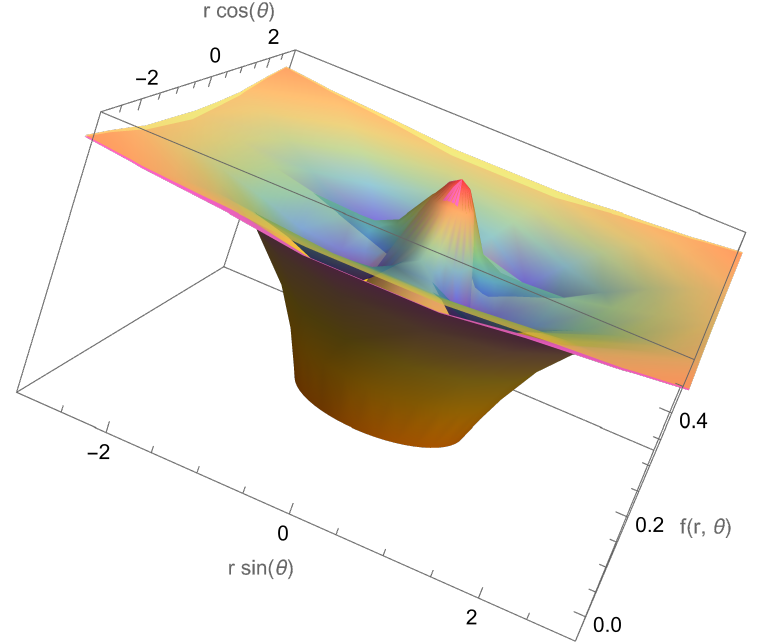}
        \caption{Zoomed-out view}
        \label{fig:uncert_product_j=2_v2}
    \end{subfigure}
    \caption{Uncertainty product of the quadratures for the GPCS with $\epsilon=2\omega$ (upper surface), and magnitude of the expectation value of their commutator (lower surface).}
    \label{fig:uncert_product_GP}
\end{figure}

\subsection{Case with \texorpdfstring{$f(\mathcal{E}_j)=f(\mathcal{E}_{j+1})=0$}{f(n)=0}}
\label{subsec:commutation_relation_f_with_roots}

Given an arbitrary state of bilayer graphene, expanded in terms of the energy eigenstates as follows,
\begin{equation}
    \ket{\Psi}=a_\nu\ket{\Psi_\nu}+\sum_{n=0}^{+\infty}a_n\ket{\Psi_n},
\end{equation}
and considering the action \cref{eq:action_general_1} and \cref{eq:action_general_2} of the diagonal ladder operators, it turns out that the commutator $[\mathbb{A}^-,\mathbb{A}^+]$ fulfills
\begin{equation}
    \label{eq:commutation_relation_I}
    \left[\mathbb{A}^-,\mathbb{A}^+\right]\ket{\Psi}
    =\Big(\mathbb{1}-j\,\mathbb{P}_{j-1}
    -\mathbb{P}_{j}+(j+1)\mathbb{P}_{j+1}\Big)\ket{\Psi}.
\end{equation}

\subsubsection{Barut-Girardello coherent states in \texorpdfstring{$\mathcal{H}_{j+1}^{\infty}$}{H j+1}}
\label{subsubsec:commutation_relation_GCS}

For the BGCS of Eq.~\cref{eq:coherent_states_CI} the use of Eq.~\cref{eq:commutation_relation_I} leads to:
\begin{equation}
\label{eq:expectation_value_PCS}
    \left\langle
    \Psi_\alpha^{\,j+1}\left|
    \left[\mathbb{A}^-,\mathbb{A}^+
    \right]
    \right|\Psi_\alpha^{\,j+1}
    \right\rangle
    =1+(j+1)P_{j+1}^{\,j+1}(\alpha).
\end{equation}
From this expression, together with Eq.~\cref{eq:commutation_relation_X_A} and the fact that the BGCS saturate the inequality \cref{eq:uncertainties_product_X_P}, we get
\begin{equation}
\label{eq:uncert_product_BGCS}
    \Delta_\alpha\mathbb{X}\Delta_\alpha\mathbb{P}
    =\dfrac{1}{2}\left(1+(j+1)P_{j+1}^{\,j+1}(\alpha)\right).
\end{equation}
According to the expression \cref{eq:transition_prob_j+1} of $P_{j+1}^{\,j+1}(\alpha)$, for large amplitudes ($r \gg 1$) the uncertainty product \cref{eq:uncert_product_BGCS} tends to $1/2$, which is the value taken by the standard coherent states.

Figure \ref{fig:uncert_product_BG} shows the uncertainty relation \cref{eq:uncert_product_BGCS}, which does not depend on $\theta$ and tends to $1/2$ for $r \gg 1$. Additionally, as $j$ increases the uncertainty product around the origin, which deviates from the standard result $1/2$, increases too.

\subsubsection{Gilmore-Perelomov coherent states in \texorpdfstring{$\mathcal{H}_0^{j-1}$}{H0}}
\label{subsubsec:commutation_relation_PCS}

Unlike the BGCS, the GPCS do not necessarily saturate the uncertainty relation \cref{eq:uncertainties_product_X_P}. Thus, we must analyze each side of Eq.~\cref{eq:uncertainties_product_X_P} to know its behavior with respect to the parameters $r$ and $\theta$ characterizing the coherent states. Considering the relationship \cref{eq:commutation_relation_X_A}, as well as the commutator \cref{eq:commutation_relation_I}, it turns out that
\begin{equation}
\label{eq:commutator_GP}
    \left|\left\langle
    \Phi_\alpha^{\,j}\left|
    \left[\mathbb{X},\mathbb{P}
    \right]
    \right|\Phi_\alpha^{\,j}
    \right\rangle\right|
    =\left|1-j\tilde P_{j-1}^{\,j}(\alpha)\right|,
\end{equation}
both in the regular and critical cases. On the other hand, the uncertainty product of the quadratures for the GPCS \cref{eq:coherent_states_GP} takes the form
\begin{align}
\label{eq:product_GP}
    \Delta_\alpha\mathbb{X}\Delta_\alpha\mathbb{P}
    =\dfrac{1}{2}\Bigg(
    &\bigg[\big(1
    +2r^2\tilde P^{\,j}_{j-1}(\alpha)\big)
    \big(1-\tilde P^{\,j}_{j-1}(\alpha)\big)
    -r^2\tilde P^{\,j}_{j-2}(\alpha)\bigg]^2\\[0.5em]
    +&\bigg[2r^2\cos{2\theta}
    \Big(
    \big(1-\tilde P^{\,j}_{j-1}(\alpha)
    \big)\tilde P^{\,j}_{j-1}(\alpha) - \tilde P^{\,j}_{j-2}(\alpha)
    \Big)
    \bigg]^2
    \Bigg)^{1/2}.
\end{align}

Figure \ref{fig:uncert_product_GP} illustrates both sides of the uncertainty relation \cref{eq:uncertainties_product_X_P} (expressions \cref{eq:commutator_GP} and \cref{eq:product_GP}) for the GPCS with $\epsilon=2\omega$. From \cref{fig:uncert_product_j=2_v1} we can see that the coherent states with $r \ll 1$ tend to saturate the uncertainty product, and as $r$ increases a region of maximum difference appears, which varies with $\theta$. However, \cref{fig:uncert_product_j=2_v2} shows that for coherent states with $r \gg 1$ the uncertainty product once again tends to saturate such inequality.

\subsection{Case with \texorpdfstring{$f(\mathcal{E}_n)\neq0$}{f(n) dif 0}}
\label{subsec:commutation_relation_f_without_roots}

Considering now the action of the diagonal ladder operators onto the eigenstates of bilayer graphene, see \cref{eq:action_general_1,eq:action_general_2}, the commutation relation between $\mathbb{A}^-$ and $\mathbb{A}^+$ turns out to be
\begin{equation}
    \left[\mathbb{A}^-,\mathbb{A}^+\right]
    =\mathds{1}-\mathbb{P}_\nu,
\end{equation}
where we must take $\mathbb{P}_\nu=\mathbb{0}$ in the critical case. In particular, for the coherent states with the standard form given by Eq.~\cref{eq:coherent_states_standard_form}, $\mathbb{A}^-$ and $\mathbb{A}^+$ commute to the identity for both cases \caseCIII and \caseCIV, i.e., for any $w_0 \in [0,+\infty)$.

Given that the BGCS saturate the uncertainty product for the quadratures, taking into account the previous discussion on the commutator of the ladder operators and Eq.~\cref{eq:commutation_relation_X_A}, it turns out that
\begin{equation}
\label{eq:uncert_product_standard}
    \Delta_\alpha\mathbb{X}\Delta_\alpha\mathbb{P}
    =\dfrac{1}{2}.
\end{equation}
This looks the same as the uncertainty product for the standard coherent states. However, in this case $\mathbb{X}$ and $\mathbb{P}$ do not represent the standard position and momentum observables.

\section{Conclusions}
\label{sec:conclusions}

In this paper we have obtained in the first place the supersymmetric partners of the shifted harmonic oscillator Hamiltonian $H_0$ by implementing the second-order confluent algorithm with an arbitrary real factorization energy.

The shifted harmonic oscillator and its SUSY partners allowed us to obtain then analytic solutions for bilayer graphene (BG) in external magnetic fields within the tight-binding model. In reference \cite{Fernandez2021_BGMagneticFields} some solutions for factorization energies $\epsilon$ in the spectrum of $H_0$ were obtained using the integral version of the algorithm. In this work we have employed the same algorithm, but for an arbitrarily real factorization energy using as well the differential method. The advantages of the last procedure are that both, the analytical expressions of the solutions and the computational implementation are simpler than in the integral algorithm.

We have analyzed also the properties of the resulting spectra, such as level spacing and degeneracy. In general, the bilayer graphene spectrum becomes partially equidistant, but there are cases with a fully equidistant spectrum. In \cref{subsec:energy_levels}, we explored the case where the bilayer graphene spectrum is equidistant without degeneracy. If $\epsilon\in\text{Sp}(H_0)$ the only case where this happens is for $\epsilon=\mathcal{E}_0$. In addition, although the external magnetic field applied to bilayer graphene varies smoothly with the parameter $w_0$, due to the properties of the SUSY transformations there is a remarkably different behavior of the set of eigenstates $\{\ket{\Psi_j}, \ket{\Psi_{j+1}}, \ket{\Psi_{j+2}}, \ldots\}$ in the regular and critical case for a given $\epsilon \in [\mathcal{E}_j, \mathcal{E}_{j+1})$, even if the spectra are identical. These differences are somehow inherited by the bilayer graphene coherent states.

In \cref{sec:coherent_states_bilayer_graphene} we derived new bilayer graphene coherent states, associated with the second-order confluent algorithm applied to the shifted harmonic oscillator for any real factorization energy $\epsilon$. This was achieved by introducing both diagonal and non-diagonal ladder operators, which are defined up to an arbitrary function $g$. The choice of such a function becomes crucial in the construction of the bilayer graphene coherent states. We have found that for $\epsilon = \mathcal{E}_j$ in the regular case and for $\epsilon \neq \mathcal{E}_j$ in both regular and critical cases the diagonal and non-diagonal ladder operators are somehow equivalent. In the first case they fail to connect successfully all the bilayer graphene eigenstates, thus we needed to choose $g$ having roots at $\mathcal{E}_j$ and $\mathcal{E}_{j+1}$; meanwhile, in the second case the ladder operators connect appropriately all the bilayer graphene eigenstates. On the other hand, for $\epsilon \neq \mathcal{E}_j$ both kind of ladder operators are not equivalent in the critical case, since in the diagonal case they fail to connect all the bilayer graphene eigenstates while in the non-diagonal case they do correctly.

According to the roots of $g$, we have obtained as well the following:

\begin{itemize}
    \item For $g(\mathcal{E}_j)=g(\mathcal{E}_{j+1})=0$, the Barut-Girardello CS form an over-complete set in the subspace spanned by $\{\ket{\Psi_n}\}_{n=j+1}^{\infty}$ and they saturate the uncertainty product for the quadratures, which in this case is always above the standard value $1/2$.
    
    Additionally, since it is not possible to derive a Barut-Girardello coherent states family in a finite-dimensional space, we decided to construct a set of Gilmore-Perelomov-like CS in the subspace spanned by 
    $\{\ket{\Psi_n}\}_{n=0}^{j-1}$. Unlike the BGCS, the GPCS do not necessarily saturate the uncertainty product for the quadratures, thus we had to analyze both sides of such inequality separately.

    Let us stress that each of these sets of coherent states exhibits temporal stability.
    
    \item For $g(\mathcal{E}_j)\neq0$ and choosing $g(\mathcal{E}_n) = 1,\,\forall\,n\in\mathbb{N}_0$, the Barut-Girardello and Gilmore-Perelomov CS constructed from the extremal state $\ket{\Psi_0}$ have the same expression as the standard coherent states. Moreover, they form an over-complete set in the subspace $\mathcal{H}_0^{\infty}=\text{span}(\{\ket{\Psi_n}\}_{n=0}^{\infty})$, which in the critical case coincides with the full state space of bilayer graphene. Additionally, the uncertainty product for the quadratures of these states takes the standard value $1/2$.
    
    Furthermore:    
    \begin{itemize}
        \item[i)] For $\epsilon<0$ the bilayer graphene coherent states are temporally stable, while for $\epsilon\neq\mathcal{E}_j>0$ they are not.

        \item[ii)] However, for $\epsilon$ such that $2\epsilon=(p/q)\omega$, with $p,q\in\mathbb{N}$ being coprime, the bilayer graphene CS evolve cyclically with a period $\tau=2\pi q/\omega_0$ at the end of which they acquire a global phase $\varphi=\pi p$. Furthermore, since the rationals are dense in the field of real numbers, for any other $\epsilon$ there exists always an approximate period $\tau$ such that $\rho(r,\theta;x,\tau)\approx\rho(r,\theta;x,0)$.
    \end{itemize}
\end{itemize}

Likewise, we computed the fidelity as function of time for any factorization energy, which in particular helps us to analyze non-cyclic evolutions and supplies extra information about what happens over a period of cyclic evolution.

In general, the geometric phase obtained for the bilayer graphene CS has an involved expression. However, it is not surprising that it coincides with the one for the standard coherent states in the case where the bilayer graphene spectrum is equidistant and non-degenerate, and when the spectrum is non-degenerate with the only non-equidistant level being the ground state energy $E_\nu = 0$.

\section*{Acknowledgments}

The authors acknowledge the support of Secihti (México), project FORDECYT-PRONACES/61533/2020.

\appendix
\renewcommand{\thesection}{Appendix}

\section{\hspace{-4.8mm}. Cyclic evolution and geometric phases}
\label{sec:cyclic_evolution}

Given a quantum system ruled by a Hamiltonian $H(t)$, we say that the system state $\ket{\psi(t)}$ evolves cyclically if after a time period $\tau$ it returns to its original state $\ket{\psi(0)}$, i.e.,
\begin{equation}
\label{eq:cyclic_evolution}
    \ket{\psi(\tau)} = e^{i\varphi}\ket{\psi(0)},
\end{equation}
where $\varphi$ is the global phase acquired during the evolution. As shown by Aharonov and Anandan \cite{Aharonov1987_Phase}, the phase
\begin{equation}
\label{eq:geometric_phase}
    \beta = \varphi 
    + \dfrac{1}{\hbar}\int_{0}^{\tau}\braket{\psi(t)|H(t)|\psi(t)}\,\mathrm{d}t,
    \quad (\text{mod } 2\pi)
\end{equation}
does not depend neither on the details of the evolution nor on the energy of the system. Thus, it is a geometric property of the physical state space known as geometric phase \cite{Moore1991}.

Although the geometric phase is commonly studied for time-dependent Hamiltonians, non-trivial geometric phases can also be derived for time-independent Hamiltonians \cite{Seleznyova1993,Fernandez1994_GeometricPhase,Fe12}. In such a case, since $H$ commutes with the time evolution operator $U$, the geometric phase for any cyclic evolution \cref{eq:geometric_phase} becomes \footnote{In this paper we consider that the geometric phase is always defined modulo $2\pi$.}
\begin{equation}
    \beta
    = \varphi + \dfrac{\tau}{\hbar}\braket{\psi(0)|H|\psi(0)}.
\end{equation}
Expanding $\ket{\psi(0)}$ in terms of the eigenstates $\{\ket{\psi_n}\}$ of $H$,
\begin{equation}
    \ket{\psi(0)} = \sum_{n}a_n\ket{\psi_n},
    \qquad a_n = \braket{\psi_n|\psi(0)},
\end{equation}
we have that
\begin{equation}
\label{eq:beta_t_indep}
    \beta = \varphi + \dfrac{\tau}{\hbar}\sum_n \abs{a_n}^2 E_n,
\end{equation}
where $\{E_n\}$ are the energy levels of the system. In particular, it is straightforward to show that any eigenstate of a time-independent Hamiltonian evolves cyclically with $\beta = 0$. However, for other cyclic states the geometric phase will generally be non-trivial. For instance, it is well known that the geometric phase for the standard coherent states is $\beta = 2\pi|\alpha|^2$, which represents the area enclosed by the closed trajectory on phase space.

It is worth noting that the geometric phase offers a deeper understanding of the structure of the quantum state space, and supplies new perspectives in various practical applications \cite{Bohm2003_GeometricPhase}. For instance, the second-order confluent algorithm allows us to obtain external magnetic fields leading to equidistant and partially equidistant spectra for the bilayer graphene effective Hamiltonian, thus the corresponding coherent states evolve cyclically with a non-null geometric phase.


\begin{thebibliography}{10}

\bibitem{Katsnelson2012}
Mikhail~I. Katsnelson.
\newblock {\em Graphene}.
\newblock Cambridge University Press, Cambridge, 4 2012.

\bibitem{DiVincenzo1984}
D.~P. DiVincenzo and E.~J. Mele.
\newblock Self-consistent effective-mass theory for intralayer screening in graphite intercalation compounds.
\newblock {\em Physical Review B}, 29:1685--1694, 2 1984.

\bibitem{Neto2009_EPGraphene}
A.~H.~Castro Neto, F.~Guinea, N.~M.~R. Peres, K.~S. Novoselov, and A.~K. Geim.
\newblock The electronic properties of graphene.
\newblock {\em Reviews of Modern Physics}, 81:109--162, 1 2009.

\bibitem{McCann2006}
Edward McCann and Vladimir~I. Fal’ko.
\newblock Landau-level degeneracy and quantum hall effect in a graphite bilayer.
\newblock {\em Physical Review Letters}, 96:086805, 3 2006.

\bibitem{McCann2013}
Edward McCann and Mikito Koshino.
\newblock The electronic properties of bilayer graphene.
\newblock {\em Reports on Progress in Physics}, 76:056503, 5 2013.

\bibitem{Setare2008}
M.~R. Setare and O.~Hatami.
\newblock Exact solutions of the dirac equation for an electron in a magnetic field with shape invariant method.
\newblock {\em Chinese Physics Letters}, 25:3848--3851, 11 2008.

\bibitem{Kuru2009}
Ş. Kuru, J.~Negro, and L.~M. Nieto.
\newblock Exact analytic solutions for a dirac electron moving in graphene under magnetic fields.
\newblock {\em Journal of Physics: Condensed Matter}, 21:455305, 11 2009.

\bibitem{Midya2014}
Bikashkali Midya and David~J Fernández.
\newblock Dirac electron in graphene under supersymmetry generated magnetic fields.
\newblock {\em Journal of Physics A: Mathematical and Theoretical}, 47:285302, 7 2014.

\bibitem{Fernandez2004_HigherOrder}
David J.~Fernandez C. and Nicolás Fernández-García.
\newblock Higher-order supersymmetric quantum mechanics.
\newblock In {\em AIP Conference Proceedings}, pages 236--273. AIP, 2004.

\bibitem{Fernandez2019_Trends}
David J.~Fernández C.
\newblock {\em Trends in Supersymmetric Quantum Mechanics}, pages 37--68.
\newblock Springer International Publishing, 2019.

\bibitem{Fernandez2020_ElectronBilayerGraphen}
David J.~Fernández C., Juan D.~García M., and Daniel O-Campa.
\newblock Electron in bilayer graphene with magnetic fields leading to shape invariant potentials.
\newblock {\em Journal of Physics A: Mathematical and Theoretical}, 53:435202, 10 2020.

\bibitem{Fernandez2021_BGMagneticFields}
David J.~Fernández C., Juan D.~García M., and Daniel O-Campa.
\newblock Bilayer graphene in magnetic fields generated by supersymmetry.
\newblock {\em Journal of Physics A: Mathematical and Theoretical}, 54:245302, 6 2021.

\bibitem{DiazBautista2017_GraphenCoherentStates}
Erik Díaz-Bautista and David~J. Fernández.
\newblock Graphene coherent states.
\newblock {\em The European Physical Journal Plus}, 132:499, 11 2017.

\bibitem{Fernandez2020_BG_CoherentStates}
David~J. Fernández and Dennis~I. Martínez-Moreno.
\newblock Bilayer graphene coherent states.
\newblock {\em The European Physical Journal Plus}, 135:739, 9 2020.

\bibitem{MNN24}
D.I. Martínez-Moreno, J.~Negro, and L.M. Nieto.
\newblock Polar coherent states in bilayer graphene under a constant uniform magnetic field.
\newblock {\em Physics Letters A}, 494:129301, 1 2024.

\bibitem{Ba24}
F.~Bagarello.
\newblock Ladder operators with no vacuum, their coherent states, and an application to graphene.
\newblock {\em Annals of Physics}, 462:169605, 3 2024.

\bibitem{DOCR20}
E.~Díaz-Bautista, M.~Oliva-Leyva, Y.~Concha-Sánchez, and A.~Raya.
\newblock Coherent states in magnetized anisotropic 2{D} {D}irac materials.
\newblock {\em Journal of Physics A: Mathematical and Theoretical}, 53:105301, 3 2020.

\bibitem{Fernandez2022_GGCoherentStates}
David J.~Fernández C. and Daniel O-Campa.
\newblock Graphene generalized coherent states.
\newblock {\em The European Physical Journal Plus}, 137:1012, 9 2022.

\bibitem{DNN21}
E.~Díaz-Bautista, J.~Negro, and L.~M. Nieto.
\newblock Coherent states in the symmetric gauge for graphene under a constant perpendicular magnetic field.
\newblock {\em The European Physical Journal Plus}, 136:505, 5 2021.

\bibitem{OD24}
Daniel O-Campa and Erik Díaz-Bautista.
\newblock Phase-space representation of coherent states generated through {SUSY QM} for tilted anisotropic {D}irac materials.
\newblock {\em Physica Scripta}, 99:105267, 10 2024.

\bibitem{Astorga2015_IntegralDifferential}
Alonso Contreras-Astorga and Axel Schulze-Halberg.
\newblock On integral and differential representations of {J}ordan chains and the confluent supersymmetry algorithm.
\newblock {\em Journal of Physics A: Mathematical and Theoretical}, 48:315202, 8 2015.

\bibitem{Fernandez2003_ConfluentAlgorithm}
David J.~Fernandez C. and Encarnacion Salinas-Hernandez.
\newblock The confluent algorithm in second order supersymmetric quantum mechanics.
\newblock {\em J. Phys. A}, 36:2537--2543, 2003.

\bibitem{FS05}
David J.~Fernández C. and Encarnación Salinas-Hernández.
\newblock Wronskian formula for confluent second-order supersymmetric quantum mechanics.
\newblock {\em Physics Letters A}, 338:13--18, 4 2005.

\bibitem{Bermudez2012}
David Bermudez, David J.~Fernández C., and Nicolás Fernández-García.
\newblock Wronskian differential formula for confluent supersymmetric quantum mechanics.
\newblock {\em Physics Letters A}, 376:692--696, 1 2012.

\bibitem{Be2016}
David Bermudez.
\newblock Wronskian differential formula for $k$-confluent {SUSY QM}.
\newblock {\em Annals of Physics}, 364:35--52, 1 2016.

\bibitem{SY18}
Axel Schulze-Halberg and Özlem Yeşİltaş.
\newblock The generalized confluent supersymmetry algorithm: Representations and integral formulas.
\newblock {\em Journal of Mathematical Physics}, 59, 4 2018.

\bibitem{AbraSteg72}
Milton Abramowitz and Irene~A. Stegun, editors.
\newblock {\em Handbook of Mathematical Functions with Formulas, Graphs, and Mathematical Tables}.
\newblock U.S. Government Printing Office, Washington, DC, tenth printing edition, 1972.

\bibitem{Prudnikov1989_Vol1}
A.~P. Prudnikov, Yury Brychkov, and O.~I. Marichev.
\newblock {\em Integrals and Series. Vol. 1. Elementary Functions}, volume~1.
\newblock Taylor \& Francis, London, 1 1992.

\bibitem{Bermudez2014_PainleveIV}
David Bermudez, Alonso Contreras-Astorga, and David J~Fernández C.
\newblock Painlevé {IV} coherent states.
\newblock {\em Annals of Physics}, 350:615--634, 2014.

\bibitem{Aharonov1987_Phase}
Y.~Aharonov and J.~Anandan.
\newblock Phase change during a cyclic quantum evolution.
\newblock {\em Physical Review Letters}, 58:1593--1596, 4 1987.

\bibitem{Moore1991}
D.~Moore.
\newblock The calculation of nonadiabatic {B}erry phases.
\newblock {\em Physics Reports}, 210:1--43, 12 1991.

\bibitem{Seleznyova1993}
A.~N. Seleznyova.
\newblock Cyclic states, {B}erry phases and the {S}chr{ö}dinger operator.
\newblock {\em Journal of Physics A: Mathematical and General}, 26:981--1000, 2 1993.

\bibitem{Fernandez1994_GeometricPhase}
David J.~Fernández C.
\newblock Geometric phases and {M}ielnik's evolution loops.
\newblock {\em International Journal of Theoretical Physics}, 33:2037--2047, 1994.

\bibitem{Fe12}
David J.~Fern\'andez C.
\newblock {Harmonic oscillator SUSY partners and evolution loops}.
\newblock {\em SIGMA}, 8:041, 2012.

\bibitem{Bohm2003_GeometricPhase}
Arno Bohm, Ali Mostafazadeh, Hiroyasu Koizumi, Qian Niu, and Joseph Zwanziger.
\newblock {\em The geometric phase in quantum systems}.
\newblock Springer, New York, 2003.

\end{thebibliography}
\end{document}